\documentclass[journal=jacsat,manuscript=article]{achemso} 

\usepackage[version=4]{mhchem}%

\author{Kim H. Pham}
\affiliation{Division of Chemistry and Chemical Engineering, California Institute of Technology, Pasadena, USA}
\author{Vijaya Begum-Hudde}
\affiliation{Department of Materials Science and Engineering, University of Illinois at Urbana-Champaign, Urbana, USA}
\author{Amy K. Lin}
\affiliation{Division of Chemistry and Chemical Engineering, California Institute of Technology, Pasadena, USA}
\author{Natan A. Spear}
\affiliation{Department of Applied Physics and Materials Science, California Institute of Technology, Pasadena, USA}
\author{Jackson McClellan}
\affiliation{Department of Chemistry, University of California, Berkeley, USA}
\author{Michael W. Zuerch}
\affiliation{Department of Chemistry, University of California, Berkeley, USA}
\author{Andre Schleife}
\affiliation{Department of Materials Science and Engineering, University of Illinois at Urbana-Champaign, Urbana, USA}
\email{schleife@illinois.edu}
\author{Kimberly A. See}
\affiliation{Division of Chemistry and Chemical Engineering, California Institute of Technology, Pasadena, USA}
\author{Scott K. Cushing}
\affiliation{Division of Chemistry and Chemical Engineering, California Institute of Technology, Pasadena, USA}
\email{scushing@caltech.edu}

\title[Article Title]{The dynamical role of optical phonons and sub-lattice screening in a solid-state ion conductor}

\begin{document}
\abstract{Solid-state electrolytes (SSEs) require ionic conductivities that are competitive with liquid electrolytes to realize applications in all-solid state batteries. Although numerous candidate SSEs have been discovered, the underlying mechanisms enabling  superionic conduction ($>$ 1 mS cm$^{-1}$) remain elusive. In particular, the role of ultrafast lattice dynamics in mediating ion migration, which involves couplings between ions, phonons, and electrons, is rarely explored experimentally at their corresponding  timescales. To investigate the complex contributions of coupled lattice dynamics on ion migration, we modulate the charge density occupations within the crystal framework, and then measure the time-resolved change in impedance on picosecond timescales for a candidate SSE, \ce{Li_{0.5}La_{0.5}TiO3} (LLTO). Upon perturbation, we observe enhanced ion migration at ultrafast timescales that are shorter than laser-induced heating. The respective transients match the timescales of optical and acoustic phonon vibrations, suggesting their involvement in ion migration. We further computationally evaluate the effect of a charge transfer from the O 2$p$ to Ti 3$d$ band on the electronic and physical structure of LLTO. We hypothesize that the charge transfer excitation distorts the \ce{TiO6} polyhedra by altering the local charge density occupancy of the hopping site at the migration pathway saddle point, thereby causing a reduction in the migration barrier for the \ce{Li^+} hop, as shown using Nudged Elastic Band (NEB) calculations. We rule out the contribution of photogenerated electron carriers and laser heating. Overall, our investigation introduces a new spectroscopic tool to probe fundamental ion hopping mechanisms transiently at ultrafast timescales, which has previously only been achieved in a time-averaged manner or solely via computational methods. Our proposed technique expands our capability to answer dynamical questions previously limited by incumbent spectroscopic strategies.}

\maketitle

\section*{Introduction}

Solid-state electrolytes (SSEs) must support high ionic conductivities that are ideally equal to or beyond that of incumbent liquid electrolytes for use in solid-state batteries, solid-oxide fuel cells, and more. The discovery of numerous \ce{Li^+} conductors like \ce{Li10GeP2S12} (LGPS) \cite{kamaya_lithium_2011} and related LGPS-type crystal structures \cite{kato_high-power_2016} like \ce{Li7P3S11} \cite{seino_sulphide_2014}, \ce{Li7La3Zr2O12} \cite{awaka_synthesis_2009}, \ce{Li_{6+x}P_{1-x}Ge_{x}S5I} \cite{kraft_inducing_2018}, and \ce{Li_{6-x}PS_{5-x}Br_{1+x}} \cite{wang_fast_2020} form the basis of many general design principles associated with high ionic conductivity \cite{jun_diffusion_2024, wang_design_2023,jun_lithium_2022}. However, these design principles often rely on a static or a quasi-static picture of the ionic conduction process. 

A dynamic picture that includes couplings between the crystal framework and the migrating ion, such as the vibrational movement of polyhedral orientations with the mobile ion, or the Coulombic interactions between the migrating ion and framework, must be considered along with the static picture \cite{ohno_materials_2020,muy_tuning_2018,krauskopf_comparing_2018}. For example, more polarizable anion frameworks generally exhibit weaker bonding interactions with the migrating ion, which flattens the energy landscape of the ion migration pathway \cite{kraft_influence_2017, muy_phononion_2021}. Lowering the energy of the average vibrational frequencies of the phonons  is also a key design strategy in enhancing ion migration \cite{muy_phononion_2021}. Minimizing Coulombic interactions is shown to enable high \ce{Li^+} conduction in \ce{Li6PS_{5-x}Se_{x}I} \cite{schlem_changing_2020} and \ce{Li6PS5X} \cite{kraft_influence_2017}, but in some cases, competing inductive effects can overpower the favorable flattening of the ion hopping energy landscape via “soft” frameworks \cite{krauskopf_bottleneck_2018,culver_evidence_2020}. Challenges still exist in our current understanding of how competing dynamical factors influence ion migration. 

Ultrafast spectroscopic tools have the ability to transiently resolve ion-framework couplings to reveal the role of lattice dynamics on ion migration. Ultrafast pulsed light sources can be used to photoexcite and study coupling effects from optical phonons \cite{forst_mode-selective_2015,vicario_narrow-band_2020,hu_anharmonic_2024}, acoustic phonons \cite{hu_anharmonic_2024}, and electrons \cite{maiuri_ultrafast_2020}, all of which are predicted to influence ion migration. For example, intensity-modulated photocurrent spectroscopy (IMPS) \cite{ravishankar_intensity-modulated_2019} and similar techniques are used to study enhanced ion mobility under photoexcitation in some solar-cell materials like \ce{CH3NH3PBI3} \cite{zhao_quantification_2017,gottesman_photoinduced_2015,kim_mapping_2023}, \ce{CsPbI_3} \cite{huang_introducing_2020}, and \ce{SrTiO3} \cite{viernstein_mechanism_2022}, where ion migration can be enhanced by up to five orders of magnitude due to the decomposition of the parent material to create vacancies or defects \cite{zhao_quantification_2017}. Enhanced mobility via photoexcitation is also possible without material decomposition \cite{gottesman_photoinduced_2015,hoke_reversible_2014}. More recently, such techniques have been used  to study opto-ionic effects in solid-state ion conductors. For instance, Defferriere \textit{et al.}\ investigated the role of the grain boundary space charge potential in an oxygen solid-state conductor, 3 mol \% \ce{Ce_{1-x}Gd_{x}O2}, enabling a nearly four-fold increase in grain boundary conduction at 250$^{\circ}$C\cite{defferriere_photo-enhanced_2022}. However, an ultrafast transient measurement of parameters directly correlated to ion migration at the timescales that optical phonons, acoustic phonons, and charge transfer phenomena occur is still missing. Such measurements would assign the exact dynamics involved based on the corresponding timescales.

In this work, we propose an ultrafast, time-resolved methodology to study how localized charge densities within the crystal framework couple to and enhance \ce{Li^+} migration in a well-characterized solid-state ionic conductor, \ce{Li_{0.5}La_{0.5}TiO3} (LLTO), which has an excitable band gap in the UV range. We employ the ultrafast impedance technique described in our previous work \cite{pham_laser-driven_2024} to transiently resolve the contributions from acoustic and optical phonons on picosecond to nanosecond timescales. We measure electron-lattice-ion dynamic coupling effects by perturbing those dynamics with different light excitations. We then measure the transient change in impedance from millisecond to picosecond timescales. For the photoexcitation of the O 2$p$ - Ti 3$d$ charge transfer transition, we measure decay signatures in the impedance measurements that correspond to the modification of ion hopping on the timescales of optical phonon vibrations, lasting tens of picoseconds. We compare these dynamics with those induced by indirectly exciting the acoustic phonon bath of the sample, leading to laser-induced heating. The signal takes more than a hundred picoseconds to decay, matching the longer baseline perturbation in the charge-transfer photoexcitation. Finally, we examine these photoinduced dynamics on longer timescales of milliseconds to seconds using single frequency impedance transient (SFIT) measurements.

Transiently resolving the changed impedance due to selective photoexcitation allows us to assign the role of high-energy optical phonons and heating in the ion hopping mechanism. Nudged elastic band (NEB) calculations~ \cite{Henkelman-2000} also support our experimental observations. When the barrier against \ce{Li^{+}} diffusion through LLTO is calculated via NEB in the excited state after promoting an electron from the highest occupied to the lowest unoccupied band, a reduction in the activation energy by 8 meV compared to the ground state is observed. The saddle point structure in the migration path and activation energy are also likely altered due to the modifications in the occupancy of the valence states from the O atoms upon excitation in the LLTO. We also prove that the measured effect is not attributed to the photogeneration of electronic carriers by using DC polarization methods with blocking electrodes and electrochemical impedance spectroscopy (EIS).

Our results suggest that the decrease in charge density of the O anions, which form the 4-O bottleneck through which \ce{Li^{+}} hops, leads to a decrease in Coulombic interactions. Distorted octahedral modes in oxide perovskites are also theorized to assist the ion hop and are likely at play in the form of phonons during the charge transfer \cite{li_enhancement_2015}. Our findings give insight into the dynamics of the crystal during an ion hop in LLTO, which can be translated to other \ce{Li^+} conductor design principles. Although we specifically investigate a \ce{Li^+} solid-state electrolyte, our methodology can extend to other ion-conducting systems with applications ranging from solid oxide fuel cell membranes to bio-inspired ion membranes. The developed technique presents a new method to understand the dynamical aspects of electronic screening and lattice vibrations in the ion hopping process.

 \section*{Results}

\begin{figure}[hbt!]
\centering
\includegraphics[width=0.5\linewidth]{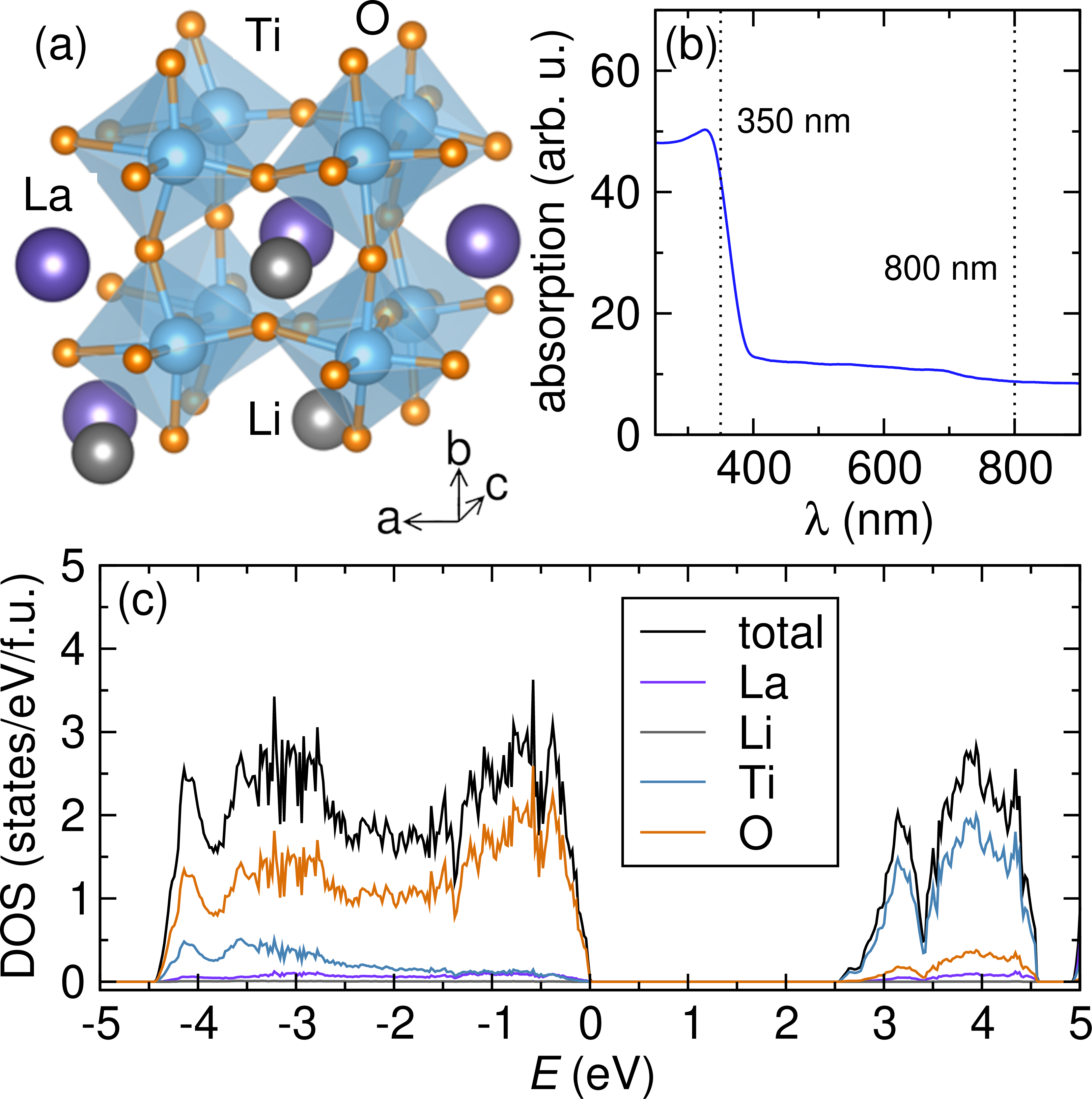}
\caption{\textbf{Crystal structure and characterization of LLTO.} (a) Crystal structure of LLTO. (b) UV-VIS spectrum of LLTO with the above band gap (350 nm) and below band gap (800 nm) wavelengths marked. The subtle feature at 700 nm is due to the internal switching from the PMT to InGaAs detector. (c) Calculated density of states (DOS) for the fully-ordered structure and projected-density of states (PDOS) associated with La, Li, Ti, and O.}
\label{fig1}
\end{figure}

LLTO is prepared via solid-state synthesis according to previous literature and is structurally characterized as reported in previous work \cite{pham_correlated_2024} and described in the methods section. LLTO crystallizes in an orthorhombic or pseudo-cubic unit cell, adopting a perovskite \ce{ABO3} crystal structure as shown in Figure \ref{fig1}a. Li and La occupy the A-site while Ti occupies the B-site. We characterize LLTO spectroscopically via UV-Vis to determine the experimental band gap energy as shown in Figure \ref{fig1}b. The band gap onset is at 380 nm (3.26 eV). The calculated density of states (DOS) for LLTO is plotted in Figure~\ref{fig1}c for the fully-ordered structure, showing that La and Li have negligible contributions to the valence and conduction band manifold in the energy range of $-$4.5 to 4.5 eV.  The O 2$p$ and Ti 3$d$ orbitals are hybridized in this energy range, indicating that Ti-O form covalent bonds. Moreover, the valence band maximum (VBM) and conduction band minimum (CBM) have predominant O-$2p$ and Ti-$3d$ character, respectively, matching previous calculations \cite{chouiekh_experimental_2023}. We note that the calculated band gap shows the typical DFT underestimation and can be improved by using a more accurate description of the electron-electron interaction, e.g., via hybrid exchange-correlation functionals or many-body perturbation theory, among other methods. For the purpose of this paper, more precise methods are not required, as we also determine the band gap experimentally.

\begin{figure}[hbt!]
\centering
\includegraphics[width=0.8\linewidth]{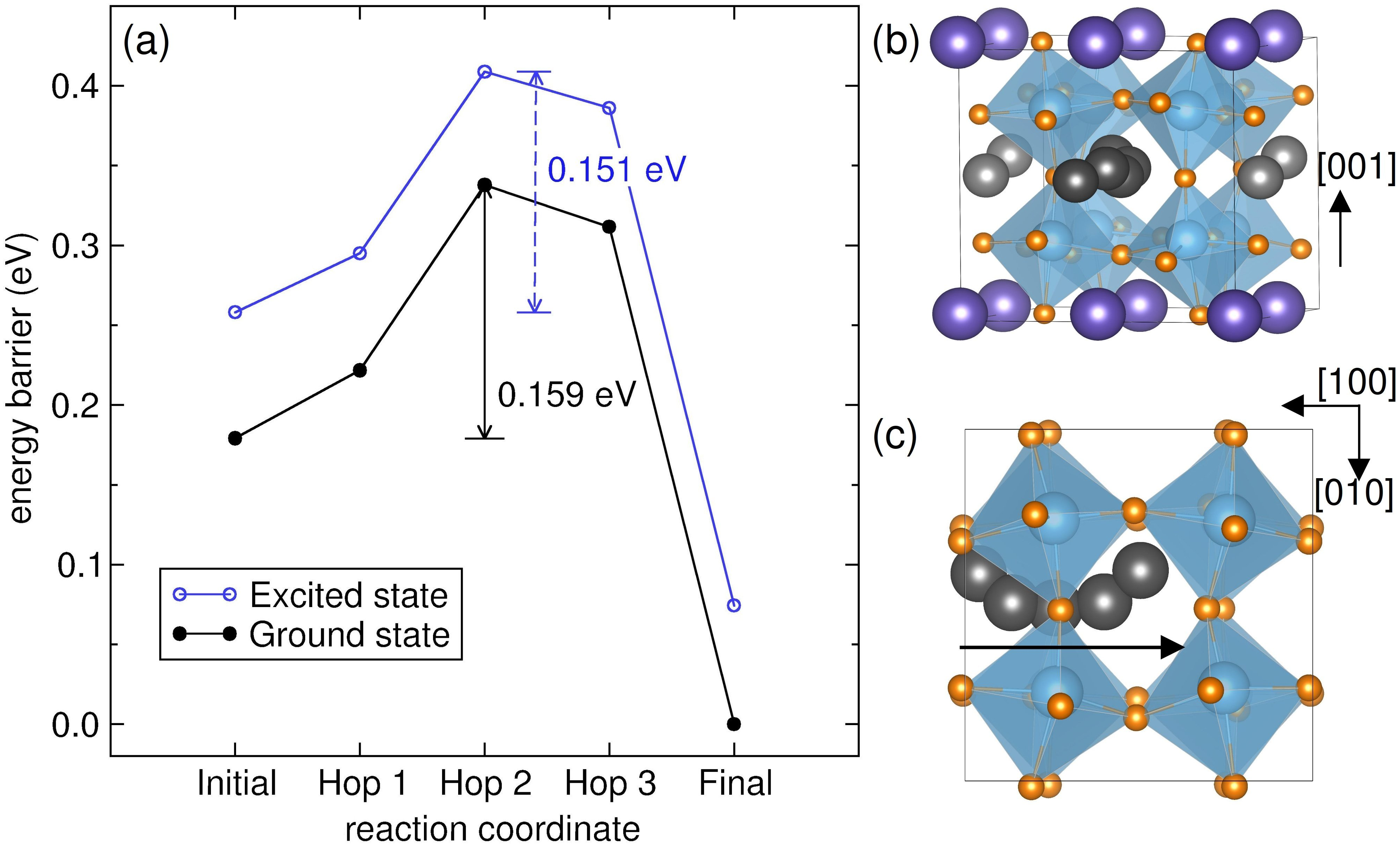}
\caption{\textbf{Calculated energy barriers of the hopping \ce{Li^+} at the ground and excited state for the fully-ordered phase.} (a) The ground and excited state migration pathways are shown as the solid black and blue lines respectively. The arrows represent the energy difference between the saddle point and the initial starting energy. 
(b) side and (c) top view of the complete NEB path, where the migrating Li atoms are marked in dark gray. In the top view, the La and non-migrating Li atoms are not shown for clarity, and the direction of migration is shown by a black arrow.}
\label{fig:theoryNEB}
\end{figure}

To predict how an above band gap excitation can influence the ion migration barrier, an electron is promoted from the valence band to the conduction band to simulate the photoexcitation. NEB calculations are first performed for the fully- and partially-ordered phase to obtain the migration barrier of \ce{Li^+} along the $a$-$b$ plane, starting from the edge of the fully-ordered LLTO structure.  For the ground state \ce{Li^+} migration in the ordered phase, a barrier of 159 meV is obtained along the NEB path (see Figure \ \ref{fig:theoryNEB}a). The positions of the migrating \ce{Li^+} ion and the NEB path in the fully-ordered structure are shown in Figure \ref{fig:theoryNEB}b-c. For the partially-ordered phase a barrier of 220 meV is obtained with NEB along the path shown in Figure\ \ref{fig:barrier2}a and Figure\ \ref{fig:barrier2}b,c, respectively. Constrained-density functional theory (CDFT) calculations coupled with NEB were performed by promoting an electron from the highest occupied band to the lowest unoccupied band at the $\Gamma-$point in both spin channels \cite{Henkelman-2000, Gali-2009}, corresponding to a carrier density of $6.9\times 10^{19}$ carriers/cm$^3$. Based on the experimental excitation parameters used in this study, we approximate carrier densities ranging between $1.9\times 10^{17}$ to $5.6\times 10^{17}$ carriers/cm$^3$ \cite{Cushing_Carriers_2018}. A reduction of the barrier of about 8 meV is observed, indicating an enhanced migration. For the partially-ordered phase, a similar reduction of the barrier was calculated.

Analysis of the band structures of the ground state geometries of the fully/partially-ordered phase along the NEB path shows a reduction in the band gap from the starting structure (2.53/2.61 eV) to the saddle point (2.26/2.30 eV), which then increases for the final structure (2.37/2.47 eV) of the NEB path (see Figure ~\ref{fig:Band1}/\ref{fig:Band2}). The band gap narrowing observed at the saddle point seems to correlate with the reduction of the migration barrier, likely due to the charge transfer altering the saddle-point crystal structure which has a non-trivial impact on the migration pathway. The saddle point of the migration pathway can be physically interpreted as the four oxygen (4-O) bottleneck window formed by four corner-sharing \ce{TiO6} octahedra as shown in Figure \ref{fig:theoryNEB}b \cite{stramare_lithium_2003}. The distortions in the \ce{TiO6} octahedra in LLTO are predicted to originate from electronic repulsions between the Ti-d and O-p orbitals in the Ti-O bond, as well as tilts and rotations driven by the disordering of the Li and La in the crystal \cite{chouiekh_experimental_2023, Kim_Gordiz_LLTO_Interplay_2024}. Therefore, the above band gap photo-excitation likely alters the degree of distortions associated with the \ce{TiO6} octahedra, thus lowering the migration barrier.

\subsection*{Ultrafast impedance measurement of ion conduction}

We use an ultrafast impedance probe to investigate the transient role of Coulombic interactions and subsequent lattice dynamics that are predicted to occur at femtosecond \cite{kim_coherent_2024} to picosecond timescales \cite{zhang_exploiting_2022,maiuri_ultrafast_2020} in LLTO. The technique is outlined in a previous publication including details on the theory and experimental methodology \cite{pham_laser-driven_2024}. In brief, the technique utilizes high-frequency electronics to measure the photo-induced change in ion migration following a femtosecond excitation at laser frequencies selected to excite the crystal structure or electronic dynamic of interest. In this study, the charge-transfer excitation from the O 2$p$ to Ti 3$d$ orbitals is investigated. Based on Figure \ref{fig1}c, pulsed 350 nm light is used, generated from a 1 kHz, 38 fs, 13 mJ Ti: Sapphire laser and Optical Parametric Amplifier. This wavelength is above the measured band gap onset of LLTO (3.26 eV). We also use 800 nm light to indirectly heat the acoustic phonon bath as a control experiment through a defect excitation route in the band gap tail. We then resolve the impedance signal using a high frequency signal generator and real-time oscilloscope, which functions similarly to a frequency response analyzer to collect EIS measurements up to gigahertz frequencies. The picosecond-resolution transients of the GHz signal are resolved after photoexcitation. 

\begin{figure}[hbt!] 
\centering
\includegraphics[width=0.5\linewidth]{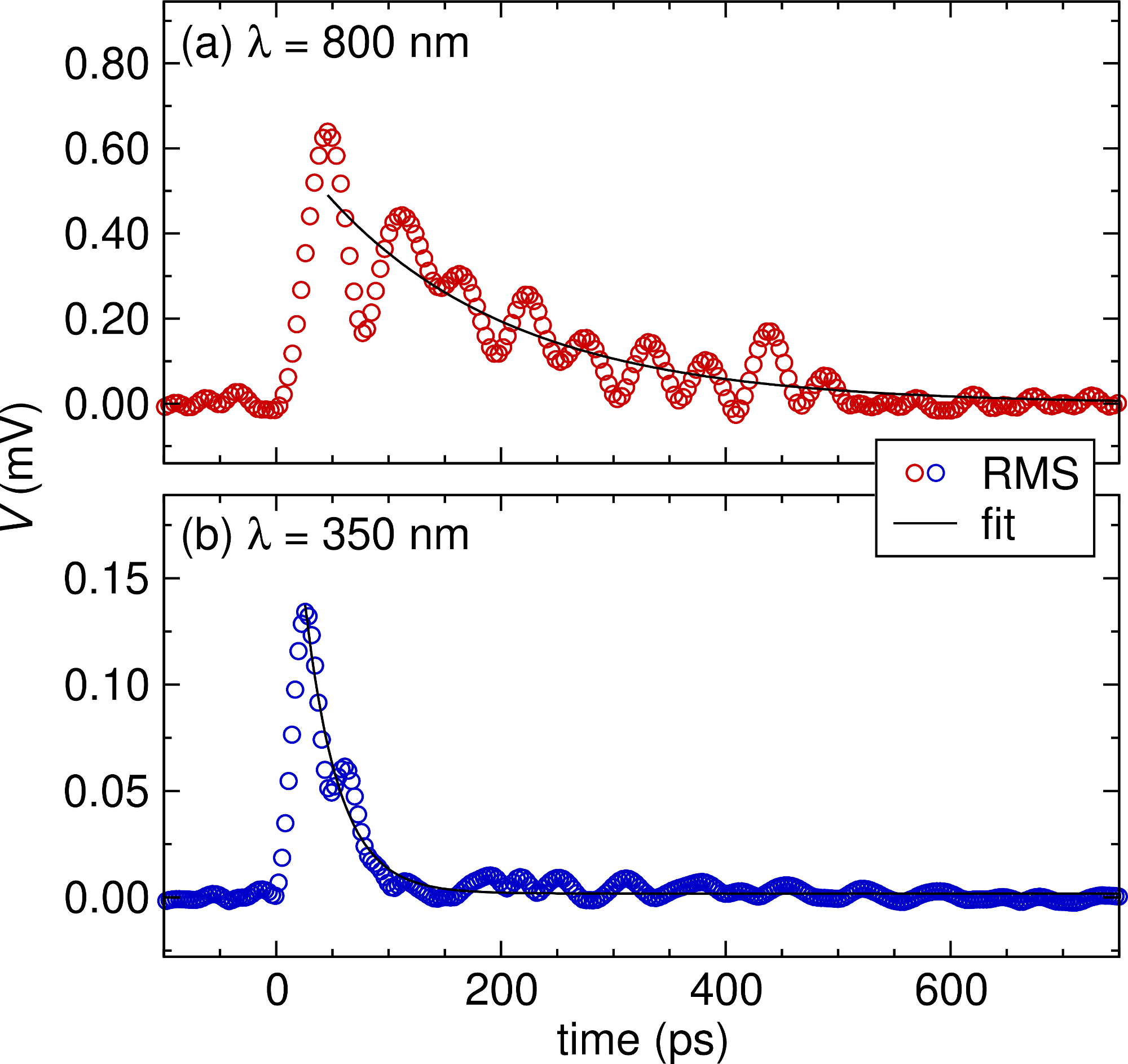}
\caption{\textbf{Comparison of the ultrafast modulation responses of LLTO from the laser-induced heating or charge-transfer excitation.} \textbf{(a)} The red and \textbf{(b)} blue circles represent the Root-Mean-Square (RMS) of the measured signal due to the 5 mW, 800 nm and 350 nm excitation, respectively, after the application of internal and external averaging, amplitude demodulation, and low pass frequency filtering. The black solid lines represent the exponential fit corresponding to the decay of the modulated signal. The input carrier frequency is at 19 GHz with an amplitude of 1 dBm. All excitations are offset to zero to compare the characteristic signal peaks and features.}
\label{fig2}
\end{figure}

The ultrafast impedance measurement is shown in Figure \ref{800nm ultrafast processing} and \ref{350nm ultrafast processing}.   The 800 nm and 350 nm excitation modulates the amplitude of a 19 GHz 1dBm carrier signal reflected off of the sample, shown in Figure \ref{800nm ultrafast processing}a and \ref{350nm ultrafast processing}a. The frequency and amplitude of the carrier signal were selected to increase the signal-to-noise ratio of the signal. We apply an amplitude demodulation function on the measured signal to extract the change in impedance. We decrease the background sinusoidal response with further averaging and a 17 GHz low pass filter, as shown in Figure \ref{800nm ultrafast processing}b and \ref{350nm ultrafast processing}b. After applying an upper root-mean-square (RMS) envelope for the processed data sets, we can accurately compare the transient impedance response of LLTO upon 800 nm laser-induced heating (Figure \ref{fig2}a) and 350 nm charge-transfer excitation (Figure \ref{fig2}b). 

The decay features of the fully processed demodulation signals are fit to an exponential decay function shown in Equation \ref{eq:1} \cite{martin_time-resolved_1994, kupfer_unravelling_2023}: 

\begin{equation} \label{eq:1}
S = A exp{\frac{-t}{\tau}}
\end{equation}

Where $S$ represents the measured signal, $A$ represents a scaling factor, $t$ represents time, and $\tau$ represents the lifetime associated with the decay. The $\tau$ is calculated from the optimized fit as shown by the black line in Figure \ref{fig2} and reported with a 95\% confidence interval. From the average of three fits, we determine $\tau$ to be 136.3 - 144.9 ps and 27.6-30.7 ps for the laser-induced heating and charge-transfer excitation signals, respectively. After the dominant exponential decay in the charge-transfer signal, the signal persists for 600 ps, consistent with the effects seen in the laser-induced heating data. We note that the instrument response function is bounded on the lower end at 7.8 picoseconds by the oscilloscope's sampling limit. Capacitive effects are likely to be present in the measurement circuit, but no signal distortion was measured up to the oscilloscope's limit of 33 GHz.  

The measured signal response from the 800 nm excitation is consistent with laser-induced heating, with the 350 nm response also eventually decaying into heat. The response from the 800 nm photoexcitation follows the known timescales of acoustic phonon excitation and decay, which would match the free carrier or defect heating mechanism present for excitation wavelengths longer than the band gap. The picosecond response triggered by the 350 nm excitation occurs on the same timescales as the decay of optical phonon vibrations following a band gap excitation, many of which are known to couple strongly to the migrating \ce{Li^+} \cite{pham_correlated_2024}. In later sections, we also conclude that photogenerated electronic effects are absent in these measurements.

\subsection*{Single frequency impedance transient measurements at millisecond to second timescales}

\begin{figure}[hbt!]
\centering
\includegraphics[width=0.5\linewidth]{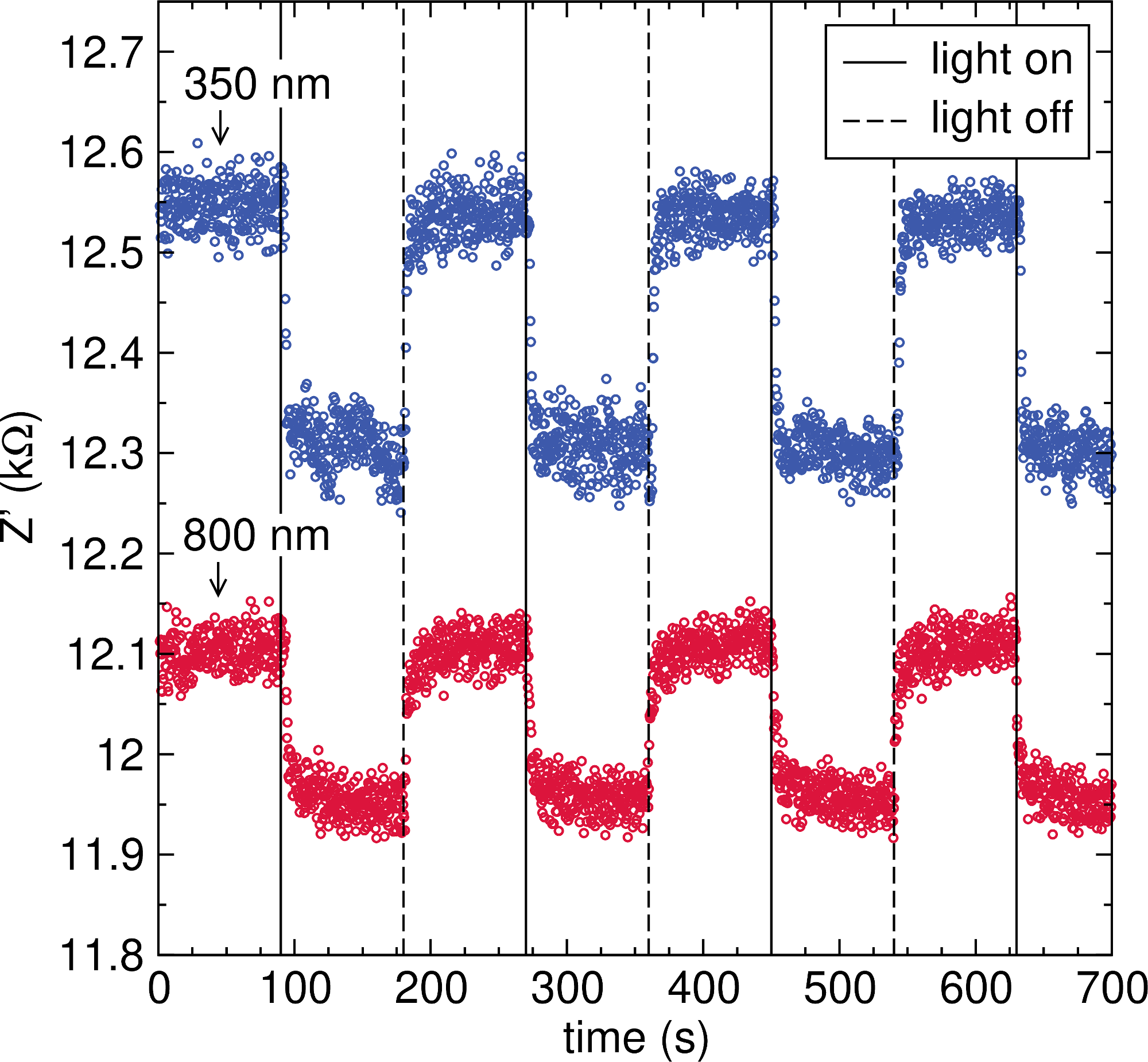}
\caption{\textbf{Single frequency impedance transients of LLTO upon 350 nm and 800 nm excitation, taken at 251 kHz}. The measurement is conducted with a frequency signal at 251 kHz and a sinus amplitude of 100 mV, corresponding to the intercept of the semi-circular feature assigned to bulk ion conduction. The solid lines indicate irradiation conditions while the dashed lines indicate the removal of the irradiation source.}
\label{fig5}
\end{figure}

We also measure the effects of photoexcitation on ion hopping at millisecond to second timescales to understand the persistence of coupled, ultrafast dynamics at timescales relevant to the practical operation of solid-state batteries. We measure impedance at a single frequency of 251 kHz with an amplitude of 100 mV, corresponding to the intercept of the bulk semi-circular feature determined first by EIS (as shown in Figure \ref{fig:350nmEIS}, with Figure \ref{fig:350nmbulksfit} showing the full change for $Z'$, $Z"$, and $|Z|$). In Figure \ref{fig5}, a single $|Z|$ value corresponding to the bulk feature is measured as a function of time with a resolution of 326 ms. The measurement is performed using chopped photoexcitation at 800 nm and 350 nm. The excitation source is repeatedly turned on and off in 90-second intervals. A higher average power is used for the 800 nm excitation source compared to the 350 nm source to match the magnitude of change in the measured $Z'$ . At 11 mW, the 800 nm excitation source shows a small decrease in $Z'$, rendering the transient analysis more difficult, as shown in Figure \ref{fig:800nmbulksfit}. The difference in required power between photoexcitation wavelengths is due to the differences in penetration depth that are wavelength-dependent, which is discussed in the final section.

For both the charge-transfer and laser-induced heating excitation, the original $Z'$ measured with the laser reproducibly returns to its initial impedance value after removal of the laser, suggesting that the observed photo-induced effect is reversible. The overall timescales for reaching steady-state impedance during the light off to on transition lasts between 9 to 14 seconds for excitation with the 800 nm source versus 6 to 8 seconds for the 350 nm source. For the opposite transition, the 350 nm impedance transient to steady-state is slower, lasting 6 to 8 seconds versus the 4 to 6 seconds in the 800 nm case. However, this conclusion is ambiguous because of the measurement resolution. More importantly, the tens of seconds required to reach steady-state between light on and off conditions is present for both excitation wavelengths, reaffirming our hypothesis that at long timescales, the decay of acoustic or optical phonons ultimately lead to heat, which can still promote ion hopping via an increase in thermal vibrations. 

Finally, we conduct a time-resolved synchrotron diffraction measurement to rule out laser-induced phase decomposition or the stabilization of a photo-induced meta-stable structural phase. The measurement is conducted on a single grain of LLTO with 100 ps resolution under 349 nm pulsed excitation. We observe that the expanded lattice constant relaxes post-illumination within 1 millisecond as shown in Figure \ref{figsynchro-us} at 302 K. This upper bound is a result of the 1 kHz repetition rate of both the 349 nm pump and X-ray probe used in the study. Since there was no overhead time between each scan, the pre-time-zero data corresponds to an effective time delay of approximately 1 millisecond between the previous pump pulse and current probe pulse. Because the observed lattice constant is equal and stagnant for each scan before time zero, the lattice must relax to equilibrium within the 1 millisecond time window between pulse events. The observed changes are associated with thermally induced lattice expansions, rather than the evolution of a new phase. We also do not measure any octahedral rotations in the transient X-ray diffraction experiments. More details regarding the time-resolved synchrotron measurement methodology can be found in previous work by McClellan and Zong \textit{et al} \cite{mcclellan_hidden_2024}. We discuss further theories to explain the measured photo-induce modulations in the Appendix.

\section*{Discussion}

The measured transients and calculated $\tau$ suggest that different dynamical processes are at play involving the migrating \ce{Li^{+}} and its interaction with the surrounding crystal. The measured response in Figure \ref{fig2} suggests that the photoexcited electron bath induced by the charge-transfer excitation thermalizes by interacting with optical phonons on a few picoseconds timescale. The laser-induced heating excitation leads to a longer decay, more consistent with the acoustic phonon bath. Acoustic phonons can also participate in electron thermalization but are more commonly observed following the decay of the optical phonon bath like that shown for $\gamma$-\ce{Li_3PO_4} \cite{hu_anharmonic_2024}. The charge-transfer excitation decay is nearly five times faster than the decay from the laser-induced heating excitation, possibly due to the modified phonons from the change in O 2$p$ charge density occupancy. In the charge-transfer transition decay signal, a smaller background signal from heat also persists on the same timescale as the acoustic phonon bath that followed the excitation of laser-induced heating, extending out to 600 ps. The matching timescales of both types of signals indicates that laser heating is present in both cases, but at a lower relative amplitude for the charge-transfer case. This charge-transfer, therefore, plays a more significant role in the modulated ionic conductivity than laser heating. 

Based on these findings, we propose the following mechanism for photo-enhanced \ce{Li^+} hopping: the above band gap, photo-induced charge-transfer excitation modifies the electronic repulsion between the Ti 3$d$ and O 2$p$ orbitals within the Ti-O bond. The modified repulsion then modulates the optical phonon bath. The changed electronic interaction distorts \ce{TiO6} octahedra, and consequentially, the 4-O bottleneck window at the saddle point of the ion migration pathway of the \ce{Li^+}. The combined effects of screening and the excited phonons lead to enhanced ion migration. 

\begin{figure}
    \centering
    \includegraphics[width=0.85\linewidth]{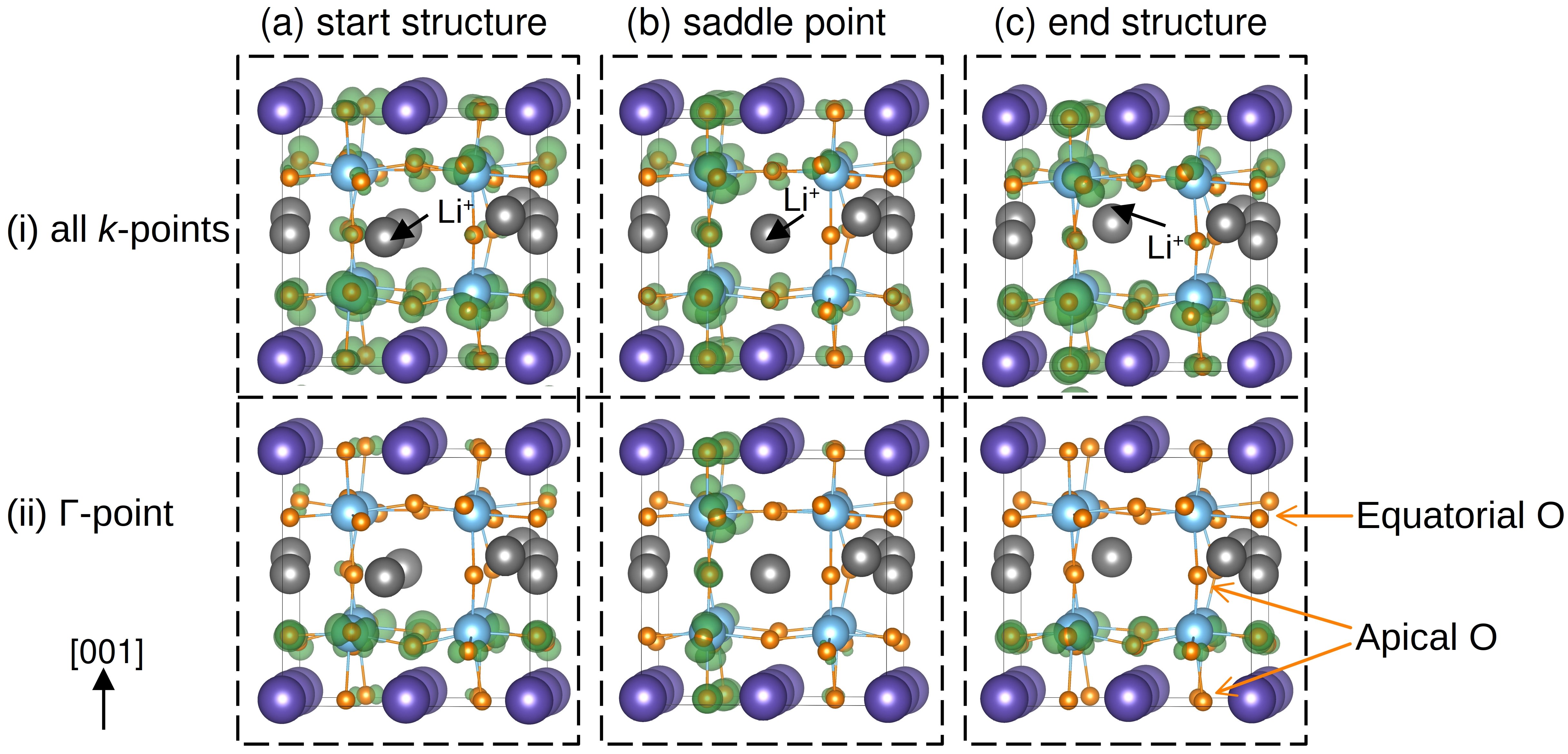}
    \caption{\textbf{Band-decomposed partial charge density plot for the fully-ordered phase of the valence band maximum (VBM).} (a) starting structure (b) saddle point and (c) end structure of the NEB path. 
    (\textrm{i}) Top row depicts charge density for all the \textit{k}-points and (\textrm{ii}) bottom row is for the $\Gamma-$point. 
    Iso-surface (in green) of the plot represents a charge density of 0.002 1/\r{A}$^3$. 
    The migrating \ce{Li^+} ion is marked with a black arrow and the NEB path is along the $a$-$b$ plane towards the centre of the structure.}
    \label{fig:Theory1}
\end{figure} 

Consistent with the hypothesis, theoretical partial charge density plots of the VBM show charge localization near the O sites, verifying their dominant contribution during optical excitation. Partial-charge-density plots for the fully-ordered phase in the ground state are presented in Figure~\ref{fig:Theory1} for the VBM and in Figure~\ref{fig:Supp-charge} for the CBM  at all \textit{k}-points in the Brillouin zone and at the $\Gamma$-point. The  O $2p$ orbitals for the VBM and Ti $3d$ for the CBM are consistent with the density of states analysis (see Figure\ \ref{fig1}c). The Li sites do not participate in the optical excitation, as shown by the absence of charge distribution around those sites. O 2\textit{p} states are observed around a majority of the O sites when the full set of \textit{k}-points in the Brillouin zone are considered. However, at the $\Gamma$ point where the excitation is initiated in the NEB calculations, the equatorial O atoms contribute significantly in the start and end structure, while predominant contributions from the apical O are observed in the saddle point structure. Notably, the O 2\textit{p} states are only observed in the \ce{TiO6} octahedra to the left of the migrating \ce{Li^+}.

Figure \ref{fig:displacement} shows the relative displacements of the relaxed excited-state geometry with respect to the relaxed ground-state geometry along the NEB path. The start and end structures reveal that the displacements of the equatorial O atoms are along the $a-b$ plane and possible twisting of the \ce{TiO6} octahedra, whereas the saddle point structure shows tilting motion of the \ce{TiO6} octahedra as seen by the displacement of the corner sharing apical O atoms mostly along the $a-b$ plane and the equatorial O atoms along the $c-$axis. We note that these displacements are observed predominantly for the \ce{TiO6} octahedra to the left of the migrating Li$^+$ and impact the Coulombic interactions in the crystal. The photoinduced charge transfer transition thus changes the screening of the crystal framework on the \ce{Li^{+}} hop, altering the migration barrier. By modifying the charge density occupancy, the structural distortions of the \ce{TiO6} octahedra are initiated, thereby changing both the 4-O bottleneck structure and the behavior of corresponding phonons, consistent with the ultrafast studies that measured a signal decay matching the timescales of optical phonons.

\subsection*{Control experiments for photo-induced thermal and electronic effects on ion mobility}

An important consideration for interpreting photo-induced enhancement in ion migration is the possibility of increased electronic conductivity or predominant thermal heating triggered by photoexcitation. First, we conduct DC polarization experiments with \ce{Li$^+$} blocking and non-blocking electrode geometries to separate the change in electronic conductivity and total conductivity under photoexcitation. Second, we conduct elevated temperature EIS measurements under photoexcitation to factor temperature into the final measured change in impedance values. Third, we use theory to model contributions from optical heating unaccounted for by our control experiment. The control experiments corroborate the conclusions discussed in the previous section.

First, we conduct DC polarization experiments with a \ce{Li^+} blocking Au electrode to discriminate between electron and ion charge carriers during photoexcitation. This experiment allows us to measure the steady-state current associated with electronic conductivity. A non-blocking Li electrode is then used to measure the total conductivity (which we verify first with EIS as shown in Figure \ref{fig:nonblockingEIS}). By subtracting the change in the electronic current from that of the total current, we can determine the enhanced current associated with \ce{Li^+} conductivity. We design and use a custom optical cell (see Appendix) to enable in-plane DC polarization measurements while exciting the sample with the above band gap light. Other experimental parameters are described in the methods section and supplementary information.

\begin{figure}[hbt!]
\centering
\includegraphics[width=0.9\linewidth]{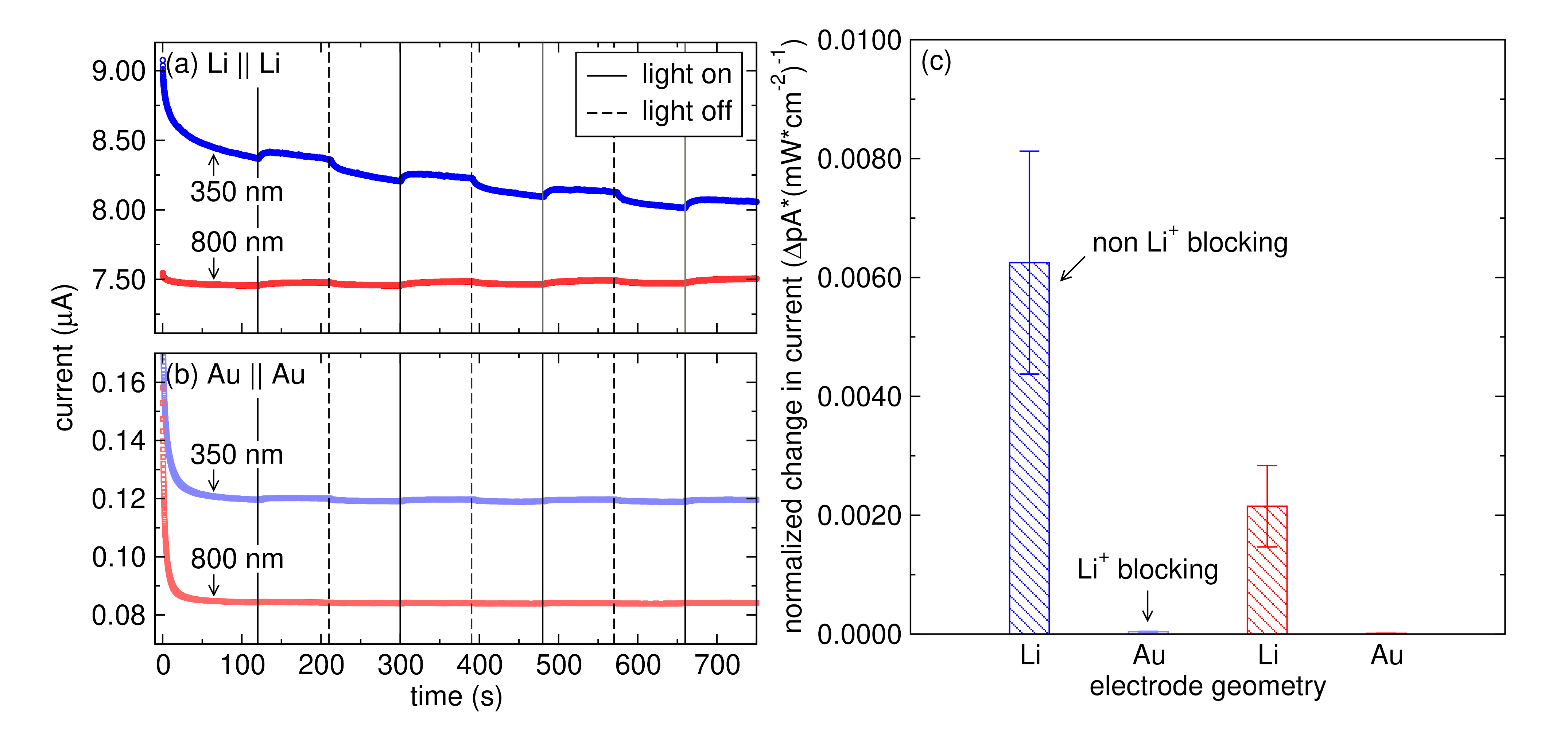}
\caption{\textbf{Steady-state current responses upon impulse 350 nm and 800 nm light with blocking and non-blocking electrodes.} Steady-state current responses after a 200 mV potential step and impulse 8 mW, 350 nm (shown as blue circles and light blue square) and 20 mW, 800 nm light (shown as red circle and light red square ) with \textbf{(a)} non-ion blocking (Li) and \textbf{(b)} ion blocking (Au) electrodes. \textbf{(c)} Change in current after normalization by power density and accounting for decreased transmission through the optical window.}
\label{DC-summary}
\end{figure}

The non-blocking electrode experiment with the 350 nm charge-transfer excitation and the 800 nm laser-induced heating excitation is shown in Figure \ref{DC-summary}a and Figures \ref{fig:350nmLiDC}-\ref{fig:800nmLiDC}. After the non-Faradaic current decay is measured, the Faradaic current associated with both \ce{Li^+} and electron mobility is measured. After 90 seconds of illumination, the 350 nm photoexcitation increases the measured current by 70-150 nA while the 800 nm heating excitation increases the current by 20-50 nA. When normalized by the power density, the charge-transfer excitation increases the total current by nearly three-fold as compared to laser-induced heating ($5.90\times 10^{-3}$ pA/mW*cm$^2$ and $2.12\times 10^{-3}$ pA/mW*cm$^2$). The ion-blocking electrode experiment is shown in Figure \ref{DC-summary}b and \ref{fig:350nmAuDC}-\ref{fig:800nmAuDC}. After the non-Faradaic current is measured, the Faradaic current associated only with electron mobility can be measured because the \ce{Li^+} does not alloy with the Au. The enhancement in current after illumination at 350 nm is 0.6-0.9 nA, which is much smaller than that measured with Li electrodes, seen in Figure \ref{DC-summary}b. Our observations are consistent with the measured ionic conductivity from the DC polarization measurements being three orders of magnitude larger than the measured electronic conductivity of $8\times 10^{-10}$ S/cm, similar to previous reports \cite{inaguma_high_1993}. Electronic carriers following the photoexcitation are, therefore, unlikely to be responsible for the measured change in impedance in the EIS measurements and the ultrafast impedance experiments.

\begin{figure}[hbt!]
\centering
\includegraphics[width=0.9\linewidth]{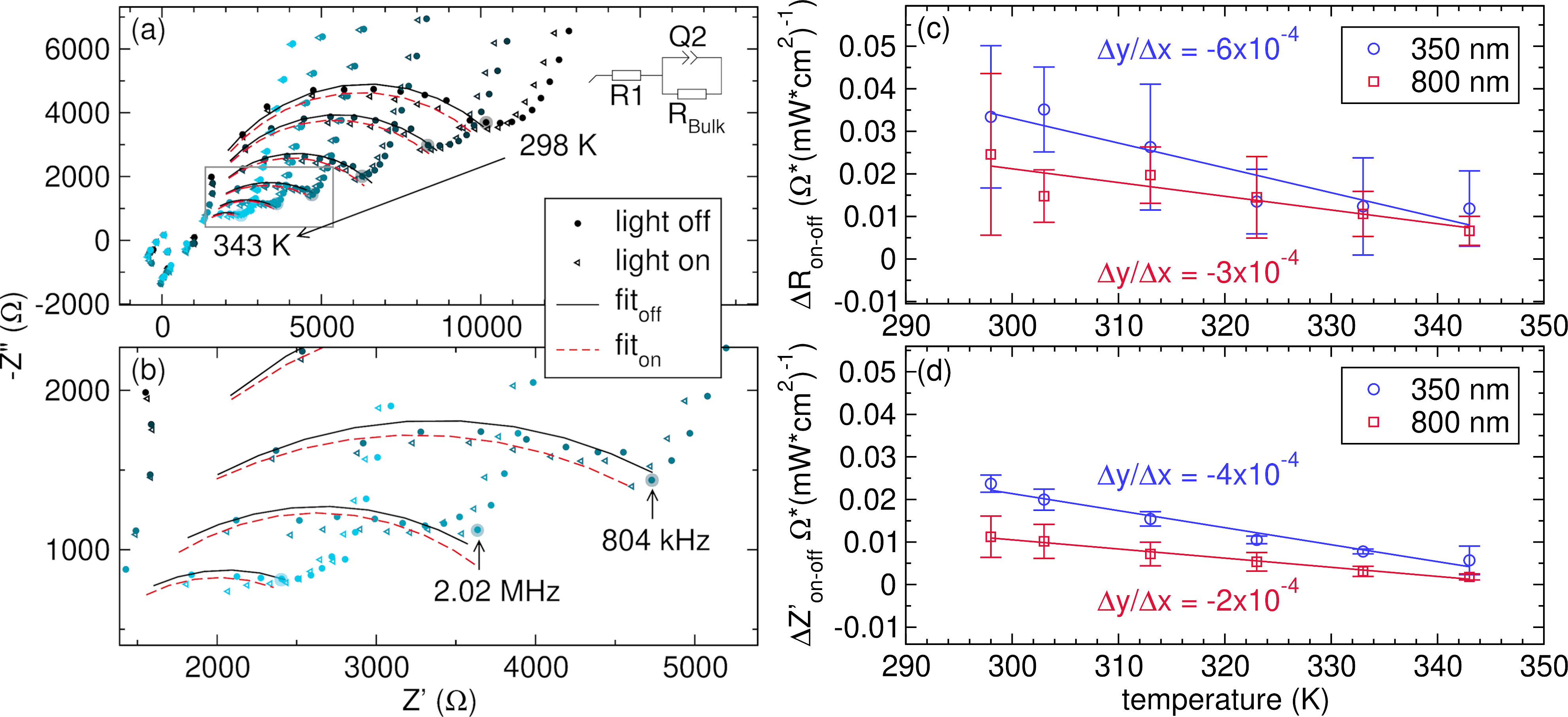}
\caption{\textbf{Nyquist plots of LLTO upon impulse 350 nm light at 7 mW and 800 nm at 20 mW at 298 K and 303 K to 343 K with 10 K increments}. \textbf{(a)} Bulk impedance semi-circle fitted to a $R1+R_\mathrm{bulk}/Q2$ circuit. The filled circles and open triangles correspond to the light off and light on measurements, respectively. The solid black lines and dashed red lines correspond to the fits for the blocked and unblocked measurements, respectively. The gray box corresponds to the region replotted in \textbf{(b)}, which highlight the poorer fits at high temperatures compared to the measurements between 298 K to 313 K \textbf{(c)}. The difference between the light off and light on measurements between 298 K to 343 K from a 350 nm or 800 nm light excitation for the fitted $R_\mathrm{bulk}$, also shown with the non-fitted intercept of the bulk semi-circle in \textbf{(d)}. The final difference is normalized by the respective power densities. $R^2$ = 0.88 and 0.78 for $\Delta R_\mathrm{bulk}$ at 350 nm and 800 nm respectively. $R^2$ = 0.97 and 0.99 for $\Delta Z'$ for 350 nm and 800 nm respectively. \textit{Reprinted from Pham et al Rev. Sci. Instrum. 95, \textbf{2024}, 073004.}}
\label{fig8}
\end{figure}

Second, we explore the contribution of laser heating to the observed enhancement of ion migration. The impedance of the sample is measured upon irradiation at temperatures between 298 K - 343 K using a custom heating cell that is described in greater detail in the Appendix. The outlined experiment enables the normalization of changed impedance by changed sample temperature. The change in impedance at elevated temperatures is shown in Figure \ref{fig8}a-b, where the Nyquist plots of LLTO are measured before (light off) and after (light on) optical excitation. The difference between the light on and light off experiments is plotted as $\Delta R$ and $\Delta Z'$ for the bulk ion migration region in Figure \ref{fig8}c and Figure \ref{fig8}d, respectively. Beyond 313 K, the impedance measurements become noisier because the higher temperatures cause the Nyquist plots to shift toward lower impedance and beyond the impedance analyzer's measurement capabilities. The resolvable semi-circular features are then buried in the noisy, high frequency region, causing poorer fits as evidenced in Figure \ref{fig8}b. The red dashed lines corresponding to the light on fit also overlap poorly with the data shown as open triangles, contributing to the uncertainty in the fit. To account for the larger standard deviation, the change in the $Z'$ intercept of the bulk semi-circular feature at 804 kHz between 298 K – 323 K and 2.02 MHz between 333 – 343 K is also plotted for comparison in  Figure \ref{fig8}d, yielding smaller standard deviations.

By calculating the slope of the measured data, the absolute change in impedance caused by the optical excitation can be normalized by temperature and power density, allowing for direct comparison between excitation wavelengths. The slope of the linear fit of the 350 nm charge-transfer excitation is $-6\times 10^{-4}$ and $-4\times 10^{-4}$ $\Omega$*(mW*cm$^{-2}$)$^{-1}$K$^{-1}$ for $\Delta R_{on-off}$ and $\Delta Z'_{on-off}$, respectively, and is double that of the 800 nm laser-induced heating excitation ($-3\times 10^{-4}$ and $-2\times 10^{-4}$ $\Omega$*(mW*cm$^{-2}$)$^{-1}$K$^{-1}$ for $\Delta R_{on-off}$ and $\Delta Z'_{on-off}$, respectively. The larger slope of the charge-transfer excitation suggests that the associated phonon bath is more sensitive to changes in ionic conduction which increases linearly with temperature  \cite{stramare_lithium_2003}.

The temperature-dependent photoexcited impedance measurements presented in Figure \ref{fig8} reveal that the magnitude of photoenhancement is dependent upon the availability of phonon baths that can be populated to enhance ion migration. Relevant phonon baths can be excited either via heating or laser-induced photomodulation. Because the 350 nm charge-transfer excitation induces a greater enhancement in ionic conduction near 298 K in Figure \ref{fig8}c-d, we believe that the phonon modes excited via this excitation couple more closely with ion transport than laser-induced heating brought about by the 800 nm pulse. More precisely, the optical phonons indicated by the ultrafast impedance measurement and excited by a charge-transfer excitation, modulate the ion hop pathway more than simply increasing thermal vibrations in the sample. If processes unrelated to ion conduction are excited, the shift in impedance would be constant and independent of temperature, yielding a slope close to zero. Additionally, if the change in impedance is purely from a heating mechanism for both the 350 nm and 800 nm excitation, the slopes would be equal when normalized by absorbed laser power density, which does not appear to be the case according to Figure \ref{fig8}c-d. While Figure \ref{fig8} indicates that the charge-transfer and heating excitation interact differently with the sample, it does not inform us about the role of thermal heating caused by a laser excitation. 

We decouple the impacts of laser heating from the bulk heating that was discussed in Figure \ref{fig8} using a finite-difference time-domain (FDTD) heat conduction model. The FDTD model, adapted from previous literature \cite{taylor_optimization_2016, taylor_integrating_2018} and further detailed in the methods section, describes the sample's change in temperature as a function of time due to laser excitation. The model is designed to account for differences in penetration depth, reflection, and additional laser power parameters when comparing the 350 nm and 800 nm photoexcitation data. The simulation results generate heating maps which show the spatial temperature variation along the cross section of the sample (Figures \ref{fig:800slice} and \ref{fig:350slice}). In the pulsed laser experiment, there is an ultrafast rise and decay in lattice temperature caused by each pulse, shown in Figures \ref{fig:800pulse} and \ref{fig:350pulse}. Within 1 second, a baseline steady state is reached as seen in Figures \ref{fig:800base} and \ref{fig:350base}. An 800 nm, 20 mW beam is predicted to increase the sample temperature by 0.5 K, while an IR thermal gun measures an increase in sample temperature between 0.4 and 0.8 K (Figure \ref{fig:irgun}). For the 350 nm, 16 mW beam, the simulation predicts an increase of 1.7 K while the measured sample temperature increase varies between 1 to 3 K (Figure \ref{fig:irgun}). 

Using the results of the model, we estimate an upper bound for the sample's increase in temperature on millisecond to second timescales. As seen in Figures \ref{fig:800slice} and \ref{fig:350slice}, the sample reaches its peak temperature along its surface. The volume probed in the EIS experiment geometry is difficult to define, but expected to extend more deeply into the sample than is directly heated by the laser. Future work will continue to explore thermal effects from the laser with improved computational models, while also utilizing electrode geometries that increase the overlap of the laser pulse with the electrochemical probe. However, these advancements are beyond the scope of this work. As such, we use the modeled surface temperature as an upper bound to estimate the laser-induced heating effects on ionic migration in LLTO.

Our results reveal that the predicted $\Delta K/mW$ of 350 nm laser power is 0.11, while the 800 nm light induces a 0.03 $\Delta K/mW$ (Figure \ref{fig:heatingmodellinregg}) increase. The difference in surface heating arises due to the two orders of magnitude shorter LLTO absorption depth at 350 nm compared to 800 nm light. We then determine the decrease in resistance attributable to thermal effects by multiplying the computed heating factor by a measured heat-induced enhancement factor of 2.16 $\% \Delta R/K$ from Figure \ref{fig8}. The final calculated thermal enhancement values, in units of $\% \Delta R/mW $, are shown in Table \ref{table1} and compared to the measured enhancements determined in Figure \ref{fig:350nmlinregg} for the 350 nm light and in previous work for the 800 nm light  \cite{pham_correlated_2024}.

\begin{table}[h!]
\centering
\label{tab:heattable}
\begin{tabular}{cccll}
\multicolumn{1}{c|}{\textbf{\begin{tabular}[c]{@{}c@{}}Wavelength\\ (nm)\end{tabular}}} & \multicolumn{1}{c|}{\textbf{\begin{tabular}[c]{@{}c@{}}Thermal Enhancement\\ (\%$\Delta$R/mW)\end{tabular}}} & \textbf{\begin{tabular}[c]{@{}c@{}}Measured Enhancement\\ (\%$\Delta$R/mW)\end{tabular}} &  & \\
\hline
\multicolumn{1}{c|}{350}                                                                & \multicolumn{1}{c|}{0.24}                                                                           & 0.37 $\pm{0.01}$                                                                           &  &  \\
\multicolumn{1}{c|}{800}                                                                & \multicolumn{1}{c|}{0.06}                                                                           & 0.10 $\pm{0.01}$                                                                            &  &  \\
\multicolumn{1}{l}{}                                                                    & \multicolumn{1}{l}{}                                                                                & \multicolumn{1}{l}{}                                                            &  & 
\end{tabular}
\caption{Deconvolution of Laser-Induced Thermal Enhancements on Ion Migration}
\label{table1}
\end{table}

From Table \ref{table1}, the total measured enhancement is larger than the contribution from thermal effects for both wavelengths. With 800 nm light, there is an excess photo-modulation effect ($0.04\; \%\Delta R/mW$), found by subtracting the thermal from the measured enhancement. The comparable values suggest that the excited acoustic phonons, which ultimately dissipate as heat, are responsible for the measured decrease in impedance. However, we make the distinction that non-resonant laser heating is not equivalent to raising the bulk sample temperature due to the presence of additional photo-modulation effects which appear to be activated, as shown by the difference in thermal and total enhancement values. In comparison, the measured change in impedance using 350 nm light is larger than the thermally-induced effects (Table \ref{table1}) by $0.13 \;\%\Delta R/mW$. Considering that the thermally-induced impact comprises 65 \% of the total measured enhancement ratio, the sample is likely heated via similar mechanisms as the 800 nm light on longer timescales. The remaining 35 \% of the total measured effects are likely attributed to above bandgap excitation effects, or the population of optical phonon baths that are inaccessible using 800 nm light. Populating the optical phonon bath leads to a greater enhancement, and is demonstrated both computationally with NEB and experimentally with the ultrafast impedance transients. Altogether, the control experiments and ultrafast impedance transients strongly indicate that the measured phenomena reflect a photo-induced change in ion migration rather than thermal or electronic effects alone.

\section*{Conclusions}

In conclusion, we conduct time-resolved impedance measurements to investigate how an above and below band gap photoexcitation enhances ion migration with picosecond resolution. We measure the role of screening and optical phonons using a charge-transfer transition. The role of 4-O phonon modes and apical oxygen charge density occupancies is explored. Our computational results support a charge transfer mechanism from the O 2$p$ to Ti 3$d$ band which leads to slight distortions in the \ce{TiO6} octahedra that form the migration bottleneck and changes in the occupied charge density surrounding the hopping \ce{Li^+}. The interplay of optical phonon modes with these dynamics leads to a reduced migration barrier and reduced impedance at ultrafast and slow timescales, as shown by NEB, EIS, and SFIT measurements. We rule out the contribution of photo-generated electronic carriers by performing DC polarization experiments with ion-blocking and non-blocking electrodes. We also explore contributions from thermal heating by conducting EIS measurements at elevated temperatures under simultaneous photoexcitation and simulating optical heating effects via FDTD methods. 

Overall, our investigation serves as an instrumental study in developing new spectroscopic tools to probe fundamental ion hopping mechanisms transiently at ultrafast timescales. We demonstrate the coupled role of optical and acoustic phonons with the migrating ion, which can be selectively excited experimentally. Our work expands upon the dynamical mechanisms of solid-state ionic conduction and motivates future work on conductive meta-stable states, memory effects, and more. Although our study only investigates one type of solid-state ion conductor, our methodology has wider applications for other types of ion and mixed solid-state conductors. 

\section*{Methods}

\textbf{NEB Calculations}
First-principles simulations were performed with the Vienna Ab-Initio Simulation Package (VASP)~ \cite{Kresse-1993,Kresse-1996,Kresse-1996-2} using the PBE exchange-correlation functional~ \cite{PBE-96}. 
We employed the projector augmented wave (PAW) method~ \cite{Blochl-1994,Kresse-1999} to describe the electron-ion interaction, using pseudopotentials for La (5$s^2$, 5$p^6$, 5$d^1$, 6$s^2$), Li (1$s^2$, 2$s^1$), Ti$_{GW}$ (3$s^2$, 3$p^6$, 3$d^4$), and O$_{GW}$ (2$s^2$, 2$p^4$).
Structural relaxations were performed with an energy cutoff of 550 eV for the plane-wave basis, employing a $4\times 4 \times 4$ and $\Gamma-$centered \textbf{k}-point mesh. 
The lattice constant was set to 7.7378 \AA~ \cite{pham_correlated_2024} and the atoms were relaxed until the forces were less than 10 meV/\AA.
For the excited state calculations, we employed the constrained DFT (CDFT) approach~ \cite{Gali-2009} with \texttt{ISMEAR=-2} flag to promote one electron from the VBM to the CBM at the $\Gamma$ point in both spin channels. 
NEB climb methodology~ \cite{Henkelman-2000} within the implementation of VASP  was employed to relax the images to compute the migration barrier.
The end structures for the minimum energy pathway were taken from a previously published work \cite{pham_correlated_2024}. 
The end structures were relaxed for both the ground- and excited-state NEB path, and three images were used to calculate the migration barriers. 
The force-based Fast Inertial Relaxation Engine (FIRE) optimizer  \cite{Henkelman-2000} was employed for the relaxations (\texttt{IBRION = 3, POTIM = 0, IOPT = 7}).
The migration barrier was calculated as the difference between the energy at the saddle point and of the starting structure of the NEB path. 

\textbf{Synthesis}
\ce{Li_{0.5}La_{0.5}TiO3} was synthesized according to literature \cite{inaguma_candidate_1994}. A stoichiometric amount of \ce{La2O3}, \ce{Li2CO3}, and \ce{TiO2} were mixed in an agate mortar and pressed into pellets under 100 MPa of pressure. The pellets were placed on a bed of sacrificial powder and calcined at 800$^{\circ}$C for 4 hours then 1200$^{\circ}$C for 12 hours at a ramp rate of 1$^{\circ}$C / minute. 

\textbf{Characterization via X-ray diffraction}
The following material was routinely characterized using a Rigaku X-ray diffractometer with CuK$\alpha$ radiation and scanned from 10 to 70 2$\theta$ at a scan rate of 0.4$^{\circ}$ per second. The following X-ray diffraction pattern was fitted using a Rietveld refinement with the GSAS II software.

\textbf{Characterization via UV-VIS}
A UV-VIS spectrum of LLTO was taken using a Shimadzu SolidSpec-3700i UV-VIS-NIR spectrophotometer over a wavelength range of 250 nm – 1000 nm in diffuse reflectance mode. The sudden decrease in absorption near 700 nm is due to the internal switching from the PMT detector to the InGaAs dectector at 870 nm. A Kubelka-Munk transform was applied to obtain the relative absorption spectrum.

\textbf{Sample preparation for EIS}
The resulting powder was pressed into a pellet with a diameter of 9 mm and a thickness of 0.8 mm under 2 tons of pressure, yielding a 75-76\% pellet density. The pellet was subsequently annealed at 1100$^{\circ}$C for 6 hours at a ramp rate of 2$^{\circ}$C / min over a bed of its mother powder. A 1.6 mm thick strip of Au was sputtered for 360 seconds at 8 W under 5 Pa onto one side of the pellet with a 1 mm gap in the center using the MSE PRO compact magnetron ion sputtering coater. The 350 mm or 800 nm beam excites the LLTO within the un-sputtered gap to eliminate possible effects on the impedance from illuminating the Au. Since the thickness of the sputtered Au electrodes are not well-defined, the accurate determination of the ionic conductivity is not possible. However, since all pellets are prepared with the same sputtering conditions, sample thickness, and diameter, the relative changes in impedance due to excitation are reproducible and reliable within error over a set of three pellets and three trials per pellet. 
To take elevated temperature EIS measurements, the sample pellet was placed inside a custom heating cell (Figure \ref{fig:heatingcell}) containing an open window for the 350 nm or 800 nm excitation source to access the un-sputtered area of the sample.

\textbf{Excitation sources}
The 350 nm and 800 nm light sources were generated using a regenerative Ti:Sapphire laser amplifier operating at 1 kHz. The 4.6 W, 800 nm output from the Ti:Sapphire laser was split using a beam splitter. One path was used directly for 800 nm excitation experiments while the other path was sent through an optical parametric amplifier and NIRUVis to generate 350 nm light at a maximum power of 12 mW - 16 mW. The power was adjusted to 2-8 mW using neutral density filters for EIS measurements under illumination. The total power of the 800 nm beam was decreased using neutral density filters to achieve 5-20 mW of average power for measurements under illumination. The beam area of the 350 nm beam and 800 nm beam is $3.76 x 10^{-4} cm^2$ and $1.06 x 10^{-3} cm^2$ respectively at their experimental maximum power. The beam size of both excitation sources were determined using a knife-edge technique. 

\textbf{Elevated-temperature EIS measurements under illumination}
The EIS spectra at room temperature were collected using the 1260A Solartron impedance analyzer with an applied sinusoidal voltage amplitude of 100 mV and a frequency range of 32 MHz to 1 Hz. 5 points are measured per decade and each point is averaged over 5 measurements. The EIS spectra at elevated temperatures from 303 K to 343 K for the grain boundary feature were collected using a sinusoidal voltage amplitude of 100 mV with a frequency range of 32 MHz to 1 Hz. 5 points are measured per decade and each point is averaged over 5 measurements. To achieve higher resolution EIS spectra at variable temperatures from 298 K to 343 K for the bulk feature, the frequency range is adjusted to a range of 32 MHz to 20 kHz and 10 points are measured per decade. The power used for the elevated temperature EIS experiments were 7 mW and 20 mW for the 350 nm and 800 nm excitation sources respectively. 

\textbf{Single frequency impedance transient measurements}
Before conducting the single frequency impedance transient measurements, an initial EIS measurement with an applied sinusoidal voltage amplitude of 100 mV and a frequency range of 1 MHz to 1 Hz was taken using a SP150 Biologic potentiostat to determine the frequency of the grain boundary and bulk intercept. The real, imaginary, and total impedance were then measured as a function of time at 10 Hz (grain boundary) and 251 kHz (bulk) frequencies with zero averaging over 700 s to obtain sub-second time resolution. The sample was measured between light and dark conditions in 90 s intervals. 

\textbf{DC polarization measurements}
To determine the electronic conductivity, an electrochemical cell matching the geometry of the heating cell, but without the thermistor, described in \ref{secA1} and shown in Figure \ref{fig:opticalcell} is prepared with Au as the ion-blocking electrode. The electronic conductivity is determined by polarizing the cell with a series of applied voltages (200, 300, 400, and 500 mV). For the non-blocking DC polarization experiment, Li foil was used in combination with a 9 mm diameter glass fiber separator sliced in half containing $30\mu$L per side of 1 M LiTFSI salt dissolved in equal volume ethylene carbonate/dimethyl carbonate to prevent the decomposition of LLTO and measure the total conductivity. To isolate the current enhancement induced by laser illumination in both the ion-blocking and non-blocking cases, the current is first measured without illumination for 120 seconds, then for the next 10.5 minutes the laser beam is alternately blocked and unblocked every 90 seconds. We use the difference between these blocked and unblocked states to determine the magnitude of current enhancement in both cases of electronic (Au electrode) and total (Li electrode) conductivity.

\textbf{Heating model simulations}
The heat accumulation and subsequent change in temperature of the samples was simulated using a finite-difference time-domain method which has previously been developed to reproduce the results of a three-dimensional two-temperature model \cite{taylor_optimization_2016, taylor_integrating_2018}. This method treats the laser pulse as instantaneous given that the time steps in the simulation are significantly longer than the length of the pulse. For this paper, 100 ns time steps were taken for a total simulation time of 1 s. A subsection of the sample with a 1.5 mm radius and 0.3 mm depth was simulated. The total simulation time was chosen because it allowed the sample to reach a clear steady state in its baseline temperature. Additionally, the time steps and total volume chosen allowed heat to effectively diffuse through the sample, while also balancing the large computational cost of simulating for long periods of time.

For this model, the thermal conductivity and heat capacity of LLTO were estimated to be similar to other oxide solid-state electrolytes \cite{wu_thermal_2022, neises_study_2022}. Given the measured porosity of our synthesized LLTO (approximately 0.264), the thermal conductivity was then adjusted using previous relations developed for other oxide materials \cite{rhee_porositythermal_1975}. While these properties fluctuate greatly with temperature, given that most laser-induced heating effects from ultrafast lasers dissipate in far less than 1 ms, the time between laser pulses, these variations are likely to be very short-lived and have minimal impact on the final temperature values.

The penetration depth of the 800 nm light was estimated to be 7 microns and the 350 nm light was estimated to be 30 nm using the following equation \cite{zong_spin-mediated_2023}:

\begin{equation} \label{penetration depth}
\delta = \frac{c}{2\kappa\omega}
\end{equation}

where $c$ is the speed of light, $\kappa$ is the imaginary part of the refractive index, and $\omega$ is the angular frequency of light. For this calculation, the imaginary part of the refractive index was taken from a previous study on the optical properties of LLTO  \cite{chouiekh_experimental_2023}.

\section{Acknowledgements}

We thank Dr. Kiarash Gordiz and Yifan Yao for helpful discussions for the NEB calculations. We thank Jax Dallas and Prof. Geoffrey Blake for fruitful discussions about the experimental work in this paper. We thank Ricardo Zarazua and Michael Roy for machining the custom optical and heating electrochemical cells at the Chemistry and Chemical Engineering machine shop of the California Institute of Technology. The solid-state UV-VIS data was collected at the Earle M. Jorgenson Laboratory of the California Institute of Technology with the assistance of Dr. Weilai Yu. We thank the Beckman Institute for their support of the X-Ray Crystallography Facility at the California Institute of Technology. The finite-difference time-domain heating model computations were conducted in the Resnick High Performance Computing Center at the Resnick Sustainability Institute at the California Institute of Technology.

\section{Funding}
KHP acknowledges funding from the National Science Foundation, Air Force Office of Science \& Research, and David \& Lucile Packard Foundation. AKL acknowledges support by the California Institute of Technology through the Beckman-Gray Graduate Fellowship. NAS acknowledges support by the U.S. Army Research Office (ARO), grant number W911NF-23-1-0001. K.A.S. acknowledges support from the Packard Fellowship for Science and Engineering, the Alfred P. Sloan Foundation, and the Camille and Henry Dreyfus Foundation. This work was supported by the U.S. Army DEVCOM ARL Army Research Office (ARO) Energy Sciences Competency, Advanced Energy Materials Program award \#W911NF-23-1-0001. The equipment used in this work was funded by the Air Force Office of Scientific Research (AFOSR), DURIP grant number FA9550-23-1-0197. VBH and AS acknowledge funding from the Illinois Materials Research Science and Engineering Center (I-MRSEC) under NSF Award Number DMR-2309037. Theoretical calculations were performed at the Illinois Campus Cluster, a computing resource that is operated by the Illinois Campus Cluster Program (ICCP) in conjunction with the National Center for Supercomputing Applications (NCSA) and which is supported by funds from the University of Illinois, Urbana–Champaign. J.M. acknowledges funding by the National Science Foundation (NSF-REU EEC-1852537). M.Z. acknowledges funding by the Department of Energy (DE=SC0024123). This research used resources of the Advanced Photon Source, A U.S. Department of Energy (DOE) Office of Science user facility operated for the DOE Office of Science by Argonne National Laboratory under Contract No. DEAC02-06CH11357.

The views and conclusions contained in this document are those of the authors and should not be interpreted as representing the official policies, either expressed or implied, of the U.S. Army or the U.S. Government.

\newpage
\section*{Supplementary information}

\renewcommand{\thefigure}{S\arabic{figure}}
\setcounter{figure}{0} 

\begin{figure}[hbt!]
\centering
\includegraphics[width=0.8\linewidth]{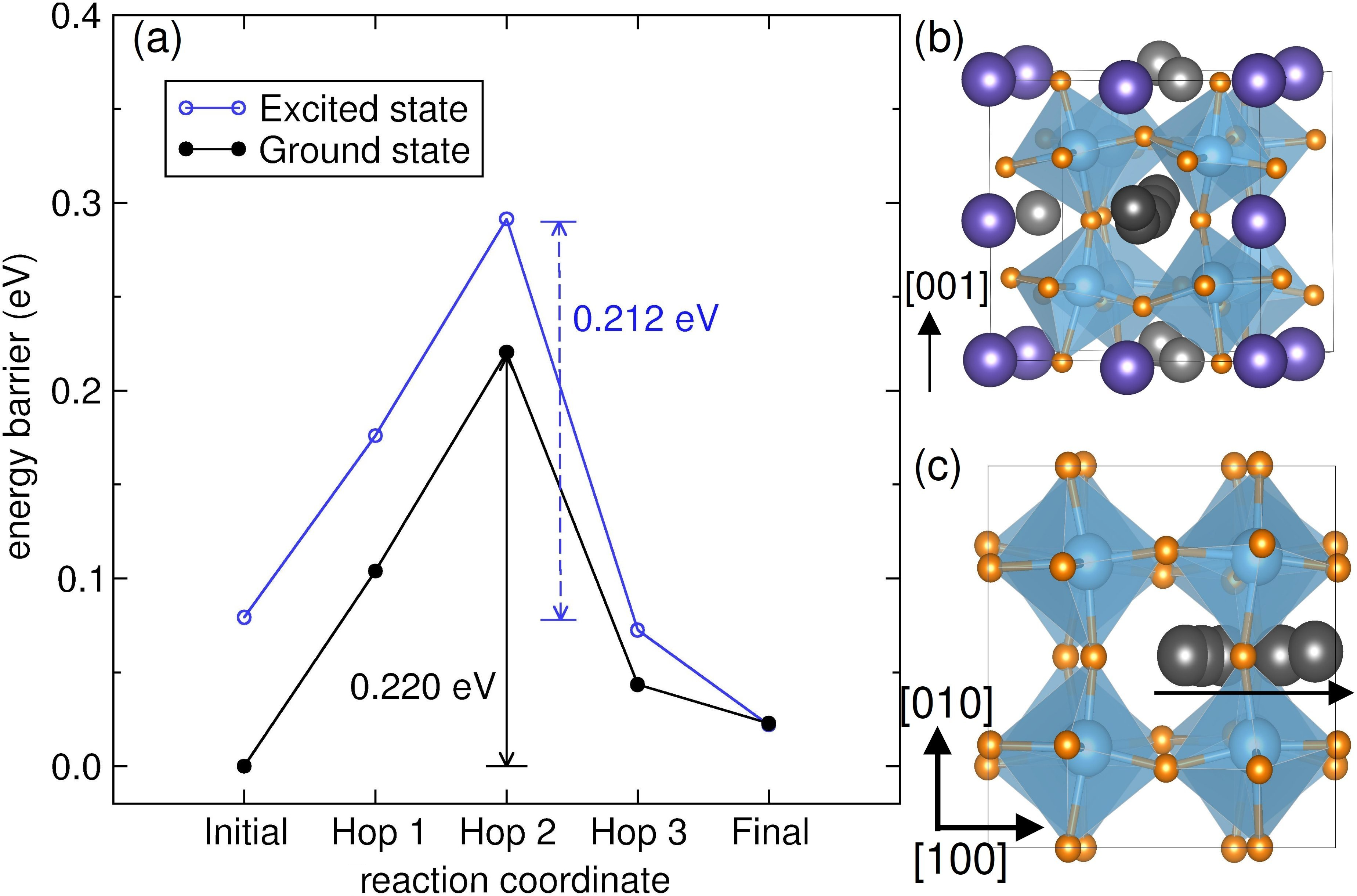}
\caption{\textbf{Calculated energy barriers of the hopping \ce{Li^+} at the ground and excited state for the partially-ordered phase.} (a) The ground and excited state migration pathways are shown as the solid black and blue lines respectively. The arrows represent the energy difference between the saddle point and the initial starting energy. 
(b) side and (c) top view of the complete NEB path, where the migrating Li atoms are marked in dark gray. In the top view, the La and non-migrating Li atoms are not shown for clarity, and the direction of migration is shown by a black arrow.}
\label{fig:barrier2}
\end{figure}

\begin{figure}[hbt!]
    \centering
    \includegraphics[width=0.95\linewidth]{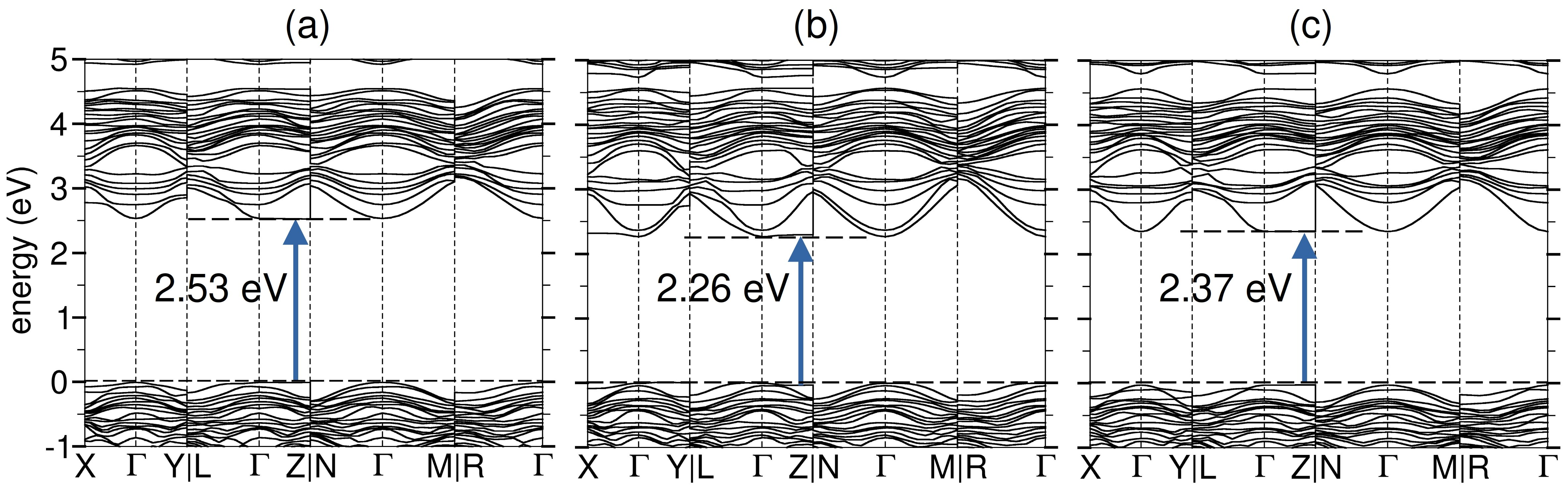}
    \caption{\textbf{Band structure of the ground state of the fully-ordered phase.} (a) starting structure of the NEB path (b) saddle point and (c) end structure of the NEB path with VBM set to zero. Band gap in (a) indirect, $\Gamma$ (VBM) to Z(CBM), (b) indirect, between $L-\Gamma$ to $\Gamma$  and (c) direct, Z to Z.}
    \label{fig:Band1}
\end{figure}

\begin{figure}
    \centering
    \includegraphics[width=0.95\linewidth]{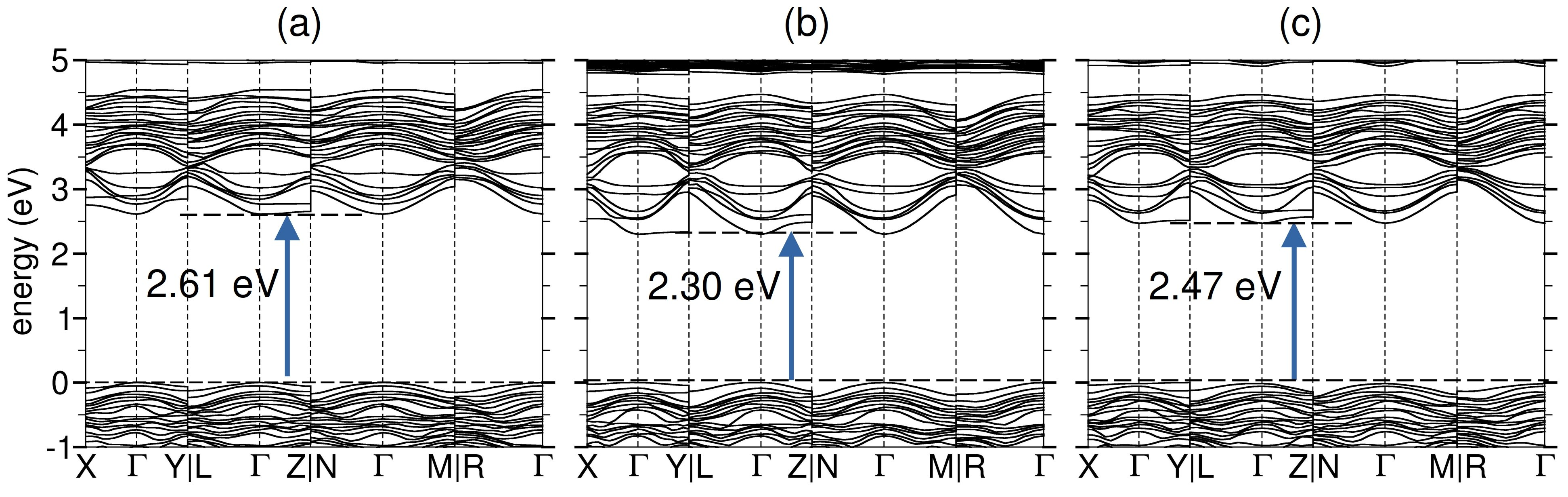}
    \caption{\textbf{Band structure of the ground state of the partially-ordered phase.} (a) starting structure of the NEB path (b) saddle point and (c) end structure of the NEB path with VBM set to zero. Band gap in (a) direct, $\Gamma$ (VBM) to $\Gamma$(CBM), (b) indirect, between $L-\Gamma$ to $\Gamma$  and (c) indirect, Y to $\Gamma.$}
    \label{fig:Band2}
\end{figure}

\begin{figure}
\centering
\includegraphics[width=0.9\linewidth]{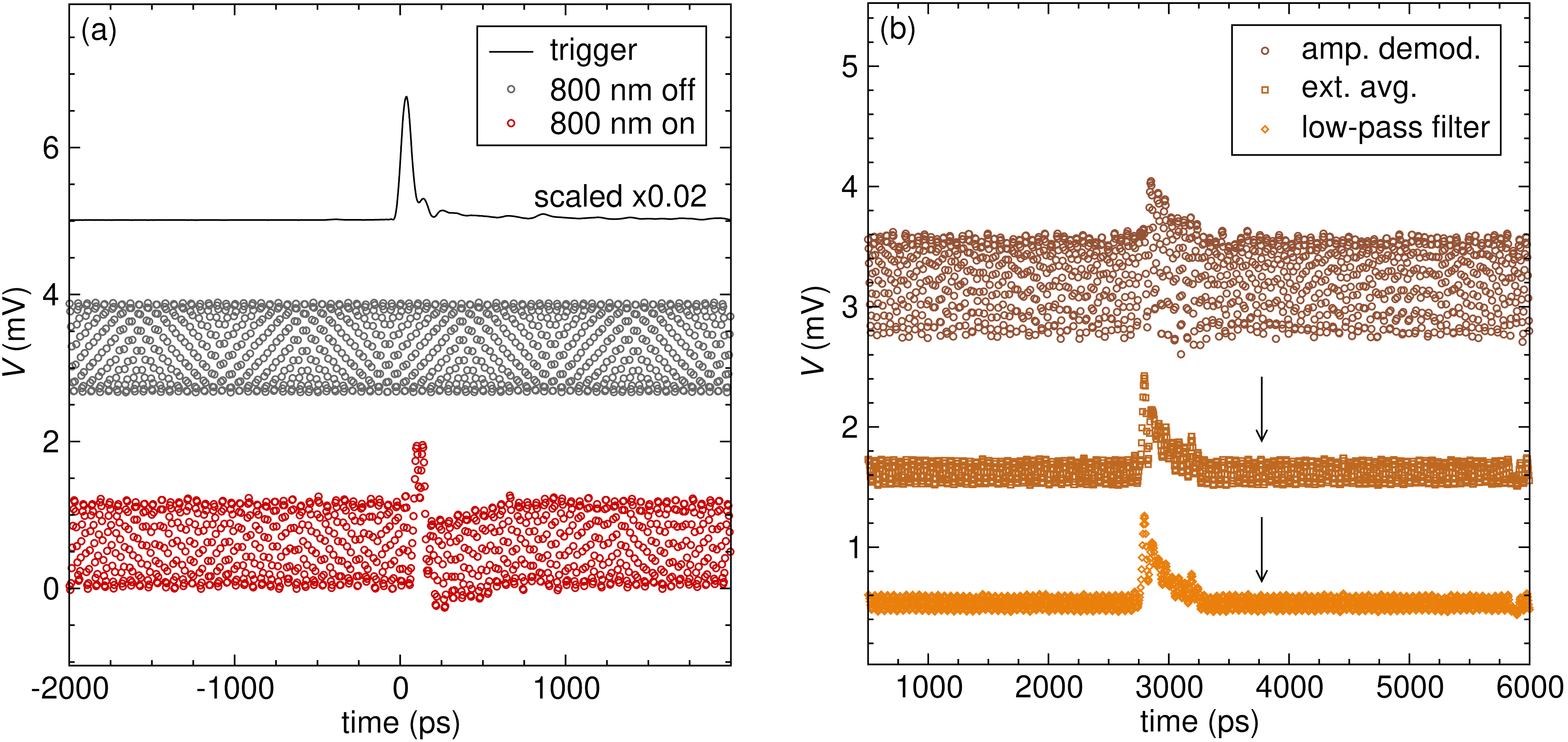}
\caption{\textbf{Measurement of the ultrafast modulated and processed amplitude demodulated signal of LLTO from 800 nm excitation}. \textbf{(a)} The measured modulated signal of the sample upon impulse 800 nm excitation at 5 mW. The black line corresponds to the 800 nm pulse used as the trigger, and the gray and red open circles represent the sample response under light off and light on conditions respectively. \textbf{(b)} The amplitude demodulated signal from \textbf{(a)} (in dark orange open circles), further processed to minimize the sinusoidal background by applying an external average (orange open squares) and a low-pass filter cut off at 17 GHz (light-orange open diamonds).}
\label{800nm ultrafast processing}
\end{figure}

\begin{figure}
\centering
\includegraphics[width=0.9\linewidth]{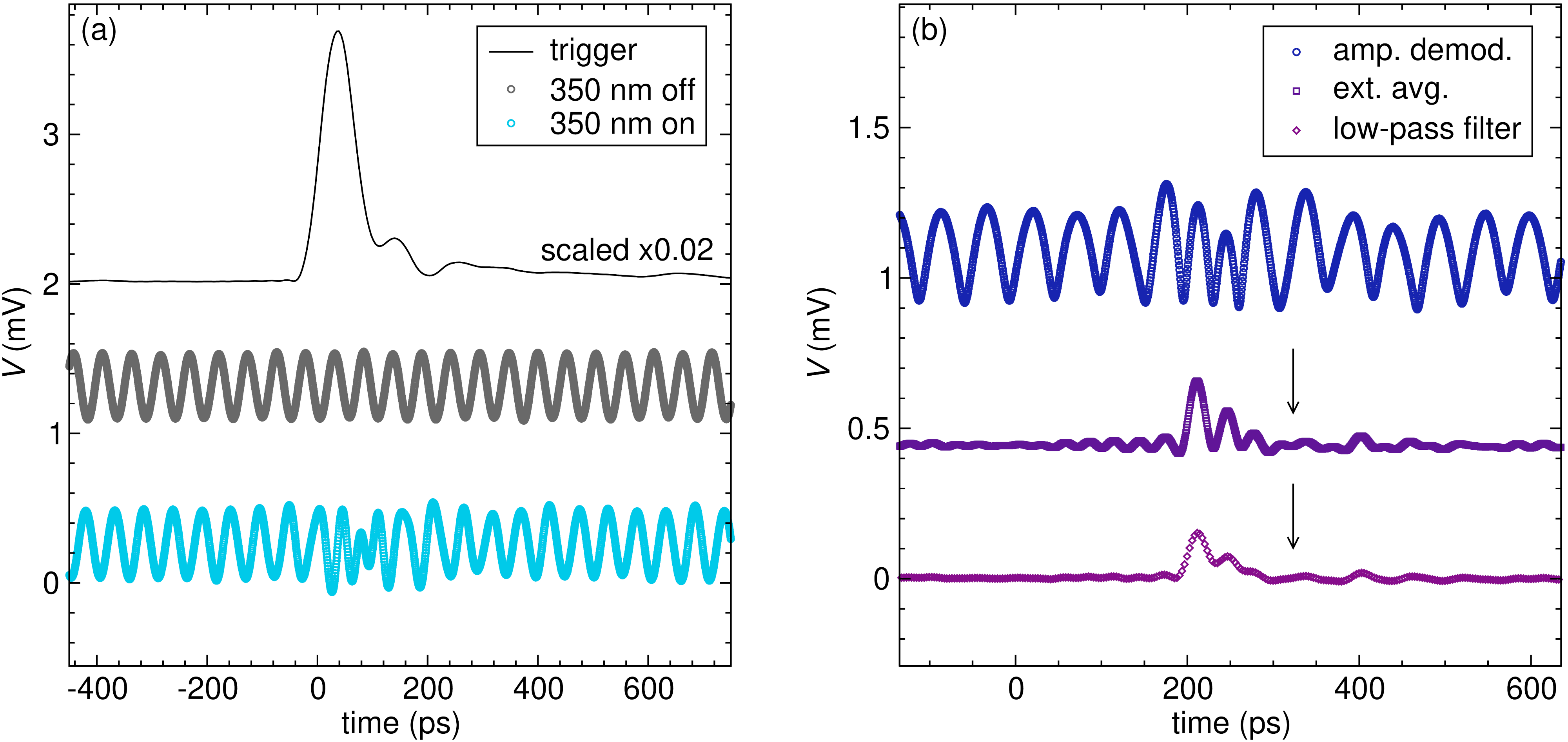}
\caption{\textbf{Measurement of the ultrafast modulated and processed amplitude demodulated signal of LLTO from 350 nm excitation}. \textbf{(a)} The measured modulated signal of the sample upon impulse 350 nm excitation at 5 mW. The black line corresponds to the 800 nm pulse used as the trigger, and the gray and blue open circles represent the sample response under light off and light on conditions respectively. \textbf{(b)} The amplitude demodulated signal from \textbf{(a)} (in blue open circles), further processed to minimize the sinusoidal background by applying an external average (purple open squares) and a low-pass filter cut off at 17 GHz (red-violet open diamonds).}
\label{350nm ultrafast processing}
\end{figure}

\begin{figure}
\centering
\includegraphics[width=0.45\linewidth]{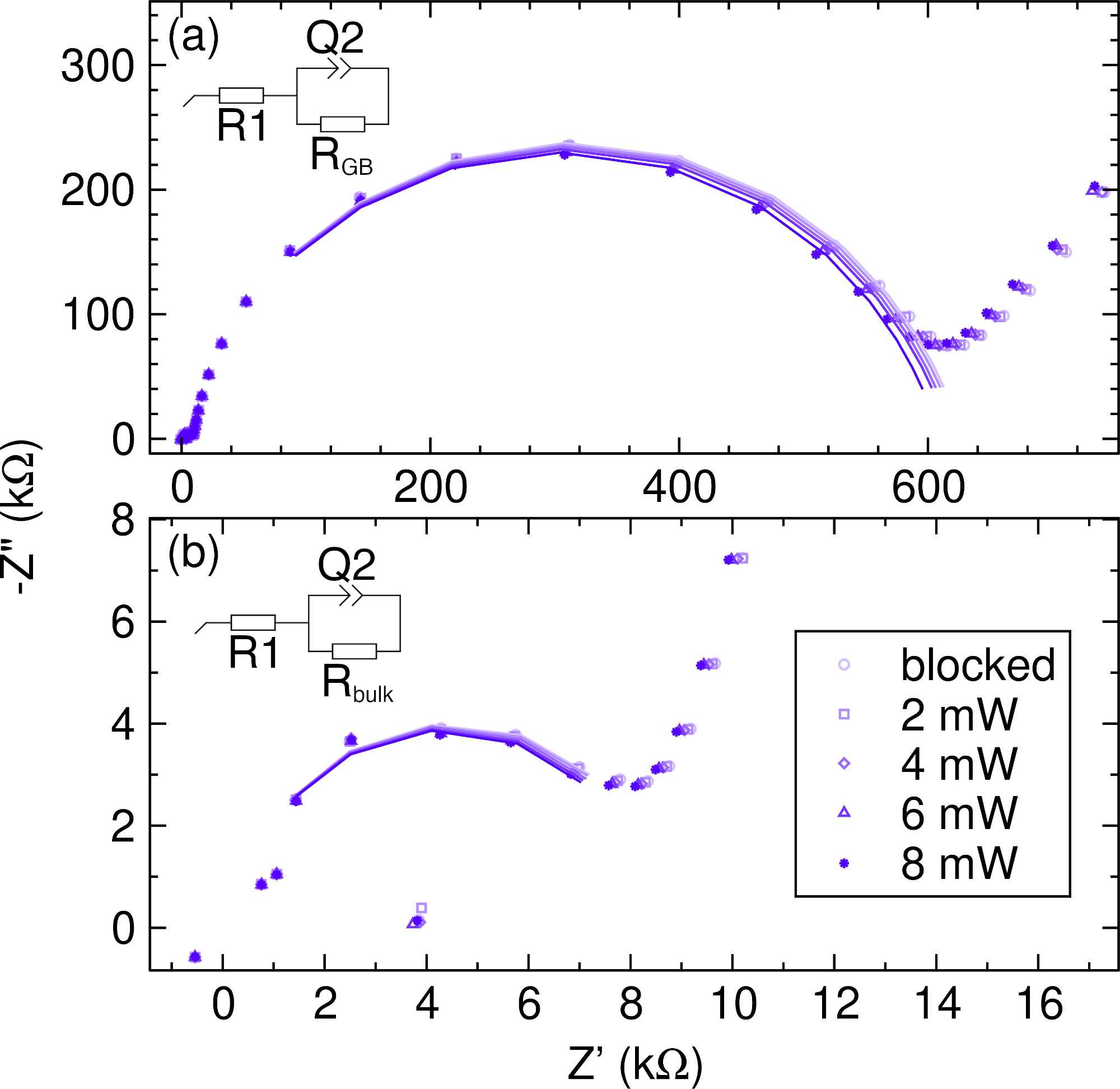}
\caption{\textbf{Reversible and linear change in the impedance} upon 350 nm excitation from 0 to 8 mW of average pulse power. \textbf{(a)} Grain boundary semi-circle fit to a R1+RGB/Q2 circuit. \textbf{(b)} Bulk impedance semi-circle fit to a R1+RBulk/Q2 circuit.}
\label{fig:350nmEIS}
\end{figure}

\begin{figure}
\centering
\includegraphics[width=0.45\linewidth]{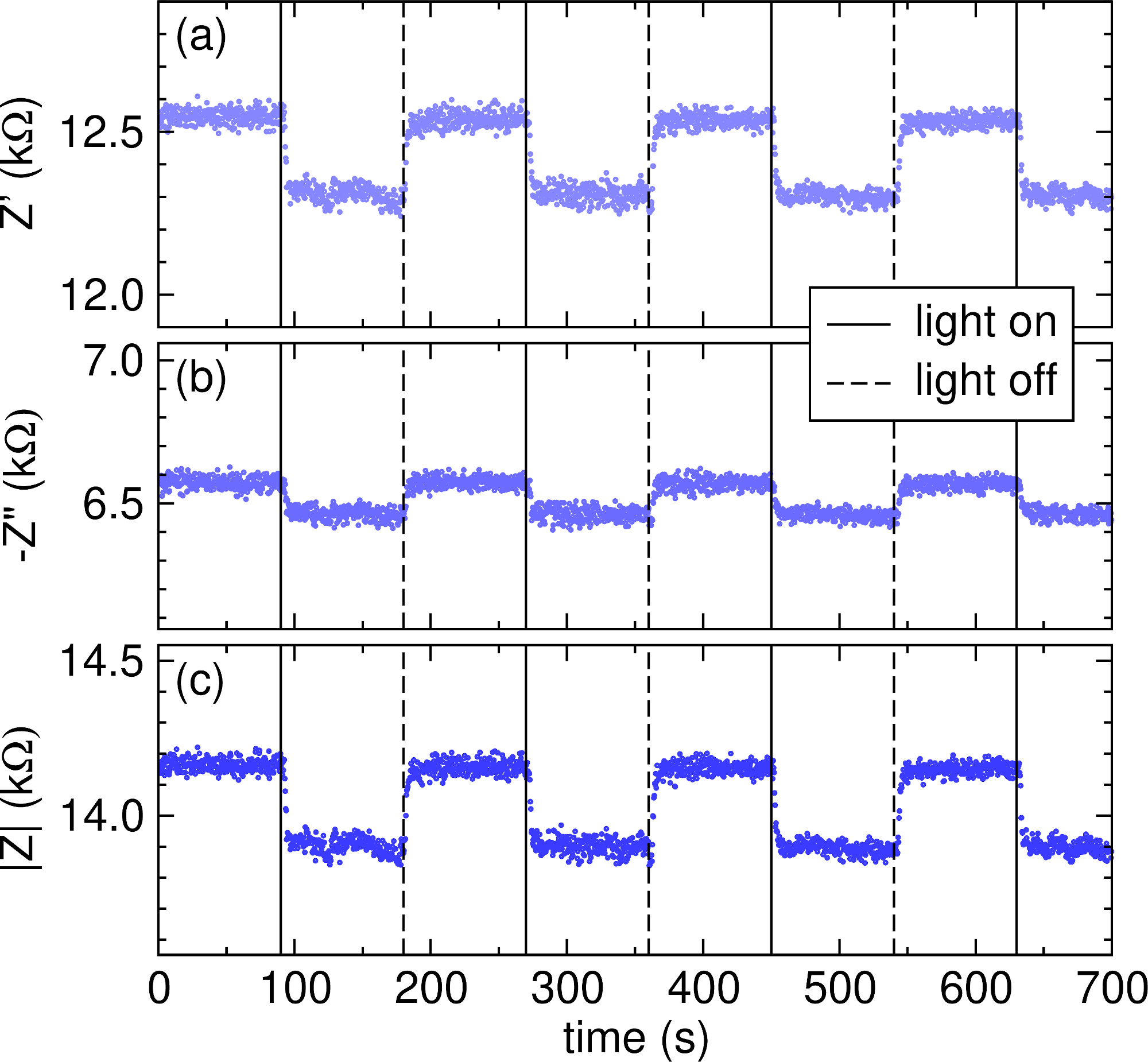}
\caption{\textbf{SFIT measurements} of (a) the real impedance, (b) the imaginary impedance, and (c) the total impedance of LLTO upon 11 mW, 350 nm excitation at 251 kHz and 100 mV sinusoidal frequency amplitude. The on and off label indicates when the impulse excitation is active or not.}
\label{fig:350nmbulksfit}
\end{figure}

\begin{figure}
\centering
\includegraphics[width=0.45\linewidth]{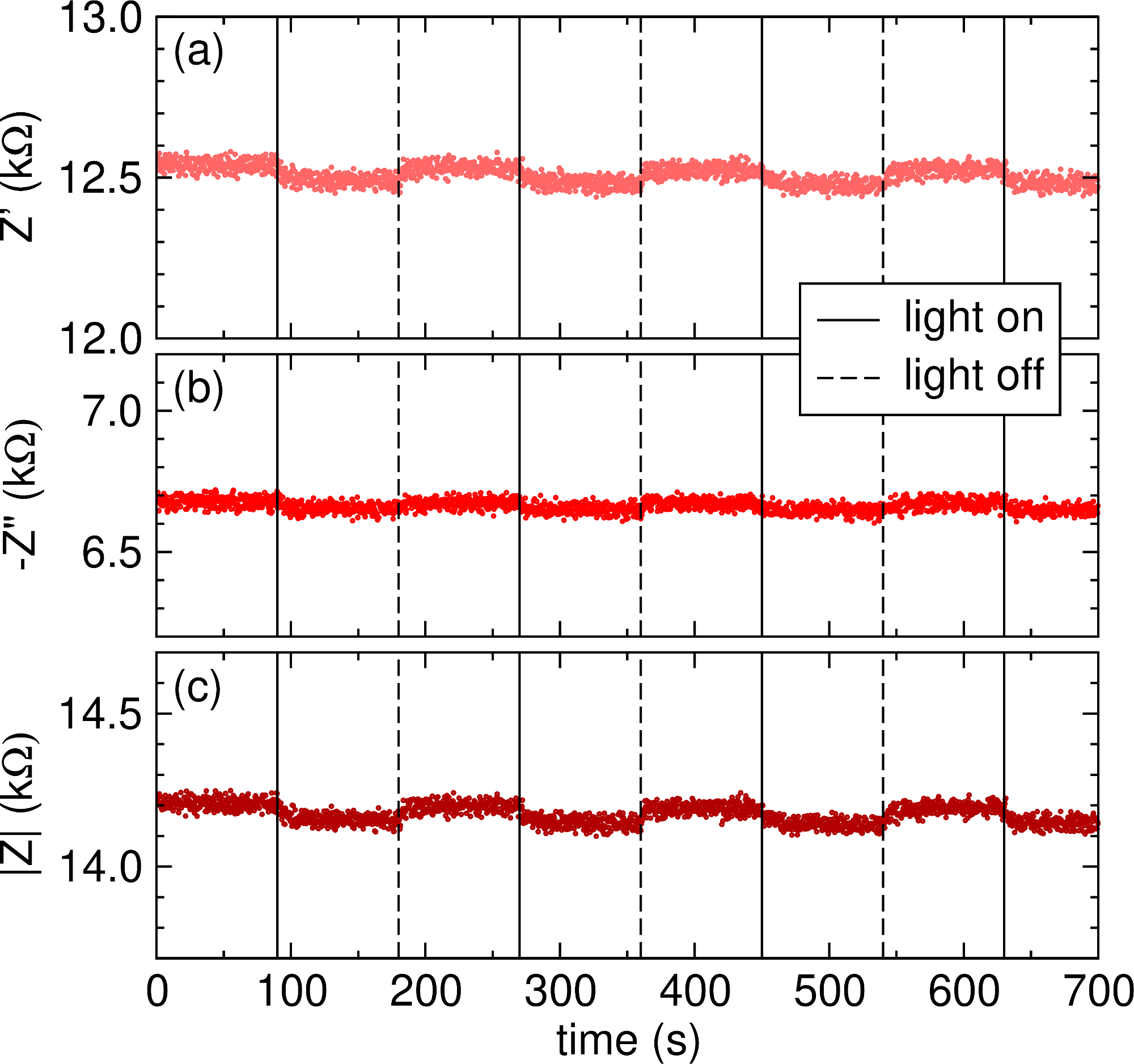}
\caption{\textbf{SFIT measurements} of (a) the real impedance, (b) the imaginary impedance, and (c) the total impedance of LLTO upon 11 mW, 800 nm excitation at 251 kHz and 100 mV sinusoidal frequency amplitude. The on and off label indicates when the impulse excitation is active or not.}
\label{fig:800nmbulksfit}
\end{figure}


\begin{figure}
\centering
\includegraphics[width=0.45\linewidth]{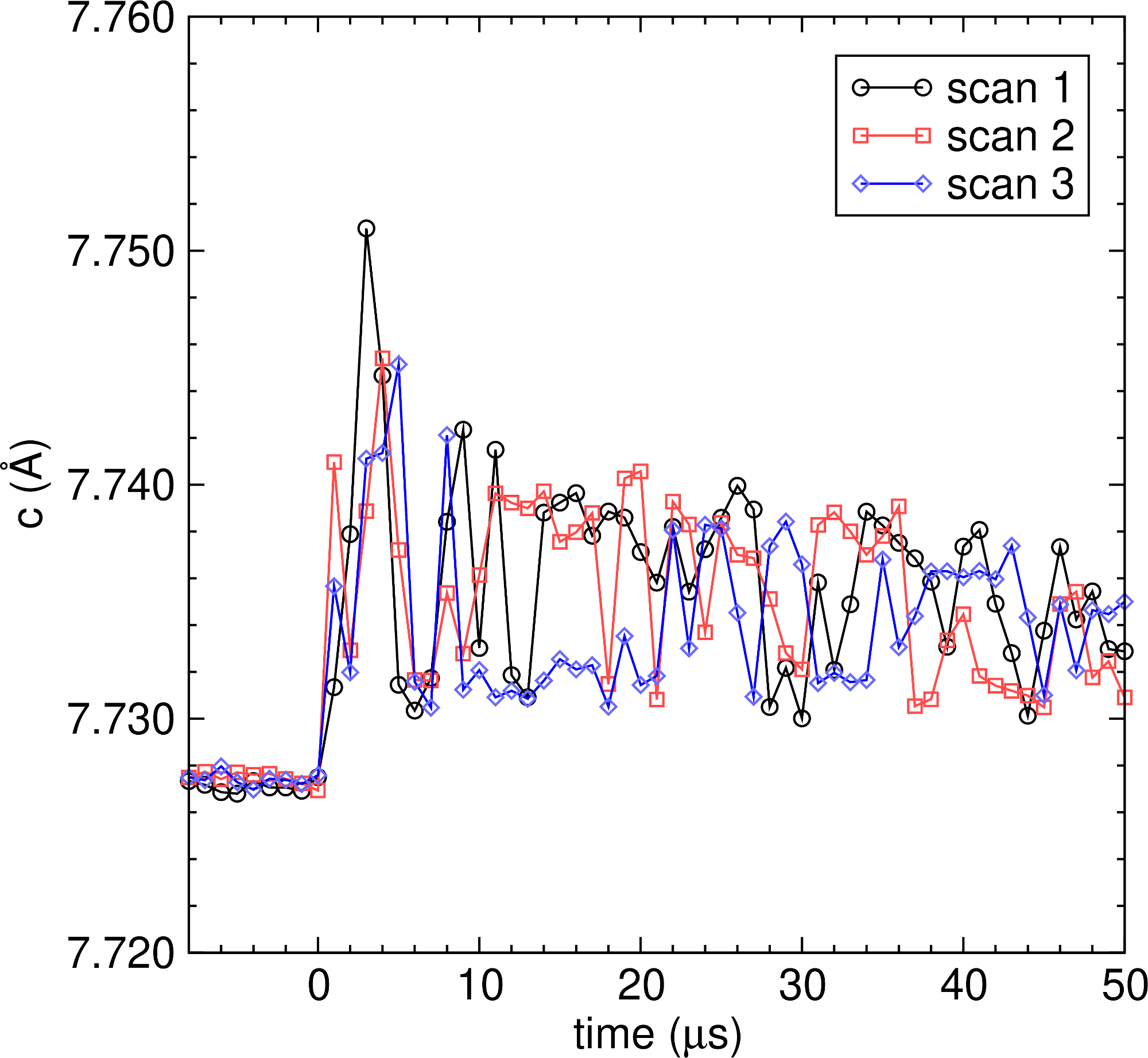}
\caption{\textbf{Time-resolved synchrotron x-ray diffraction of the c-parameter of a single grain of LLTO, upon excitation of a 349 nm, 67 $mJ/cm^2$, 1 kHz repetition rate source.} The excitation occurs after 0 $\mu s$, increasing the lattice parameter initially which then eventually recovers back to the original value overtime.}
\label{figsynchro-us}
\end{figure}

\begin{figure}[hbt!]
    \centering
    \includegraphics[width=0.85\linewidth]{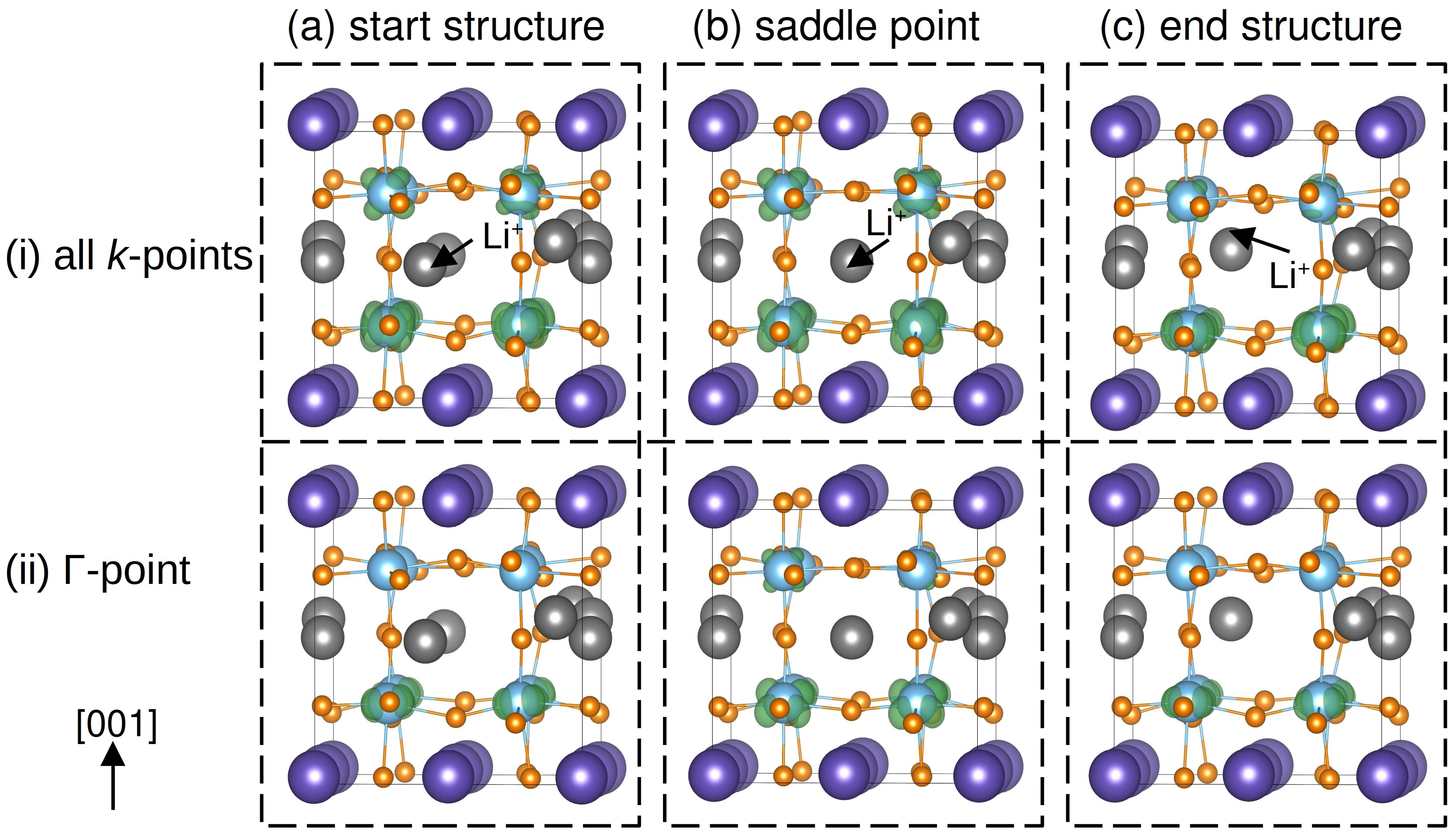}
    \caption{\textbf{Band-decomposed partial charge density plot for the fully-ordered phase of the conduction band minimum (CBM).} (a) starting structure (b) saddle point and (c) end structure of the NEB path. 
    (\textrm{i}) Top row depicts charge density for all the \textit{k}-points and (\textrm{ii}) bottom row is for the $\Gamma-$point. 
    Iso-surface (in green) of the plot  represents a charge density of 0.002 1/\r{A}$^3$. 
    The migrating \ce{Li^+} ion is marked with a black arrow and the NEB path is along the $a$-$b$ plane towards the center of the structure.}
    \label{fig:Supp-charge}
\end{figure} 

\begin{figure}[hbt!]
    \centering
    \includegraphics[width=0.9\linewidth]{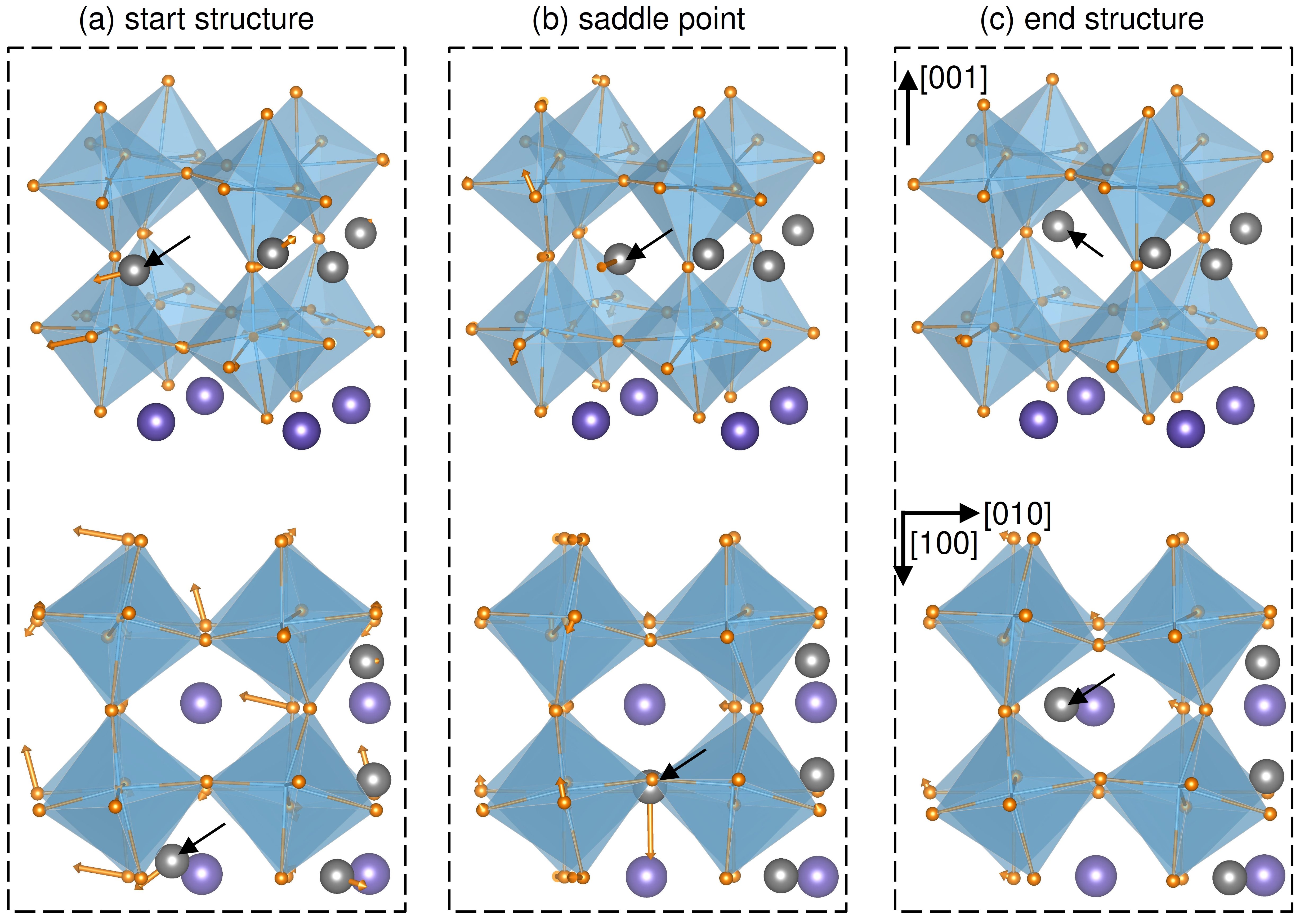}
    \caption{\textbf{Displacements in the fully-ordered phase calculated as the distance between the relaxed atomic positions of the excited and ground state structures.} 
    Orange arrows indicate the displacements for the relaxed ground state geometry (tail of the arrow) with respect to excited-state geometry (head of the arrow) in the NEB path for (a) start structure, (b) saddle point and (c) End structure. Black arrows mark the migrating Li$^{+}$ ions.}
    \label{fig:displacement}
\end{figure} 

\begin{figure}
\centering
\includegraphics[width=0.45\linewidth]{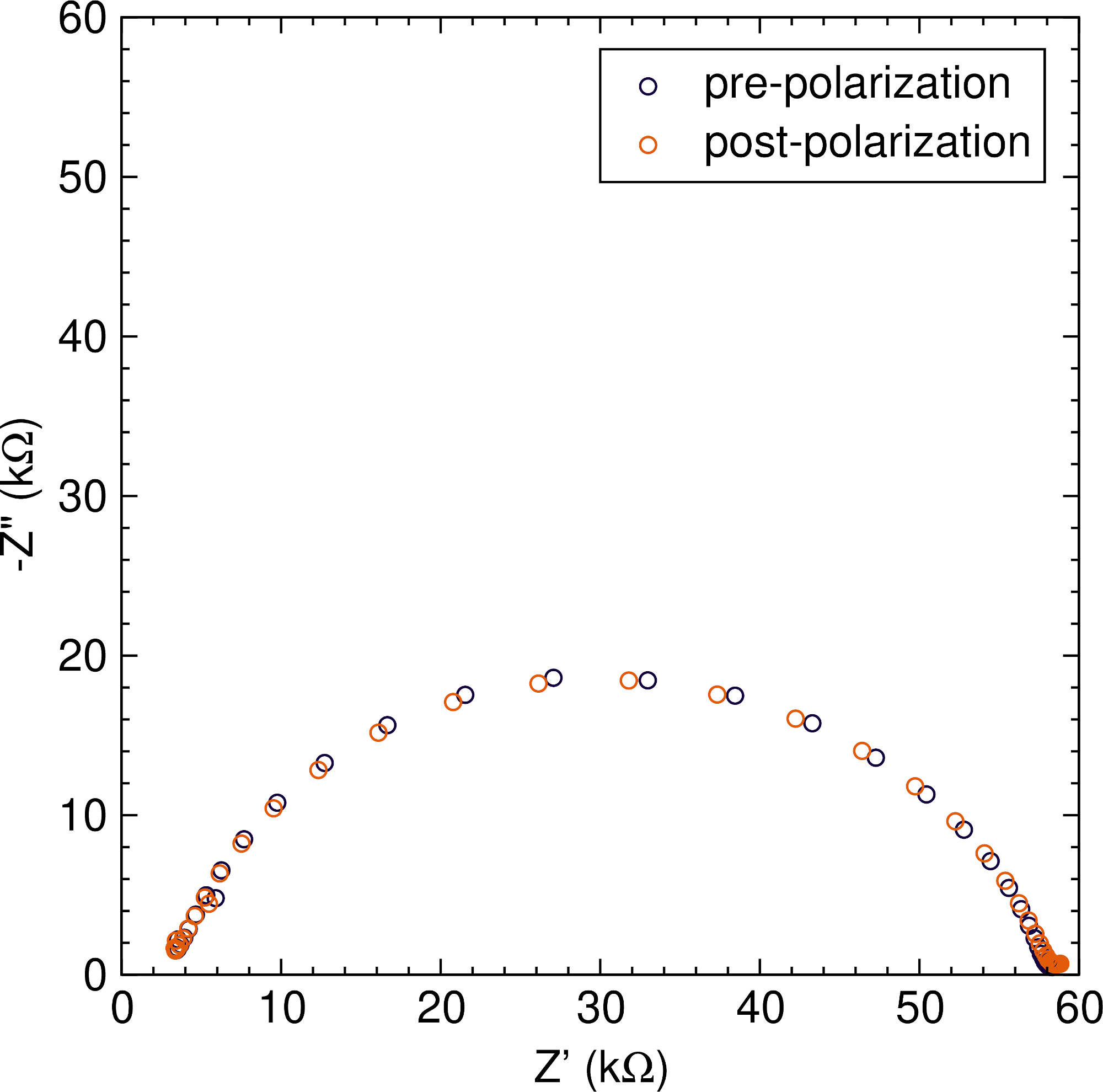}
\caption{\textbf{Nyquist plot for LLTO using non-blocking Li electrodes and 1 M \ce{LiTFSI}} before and after DC polarization measurements under 350 nm and 800 nm irradiation.}
\label{fig:nonblockingEIS}
\end{figure}

\begin{figure}[hbt!]
\centering
\includegraphics[width=0.45\linewidth]{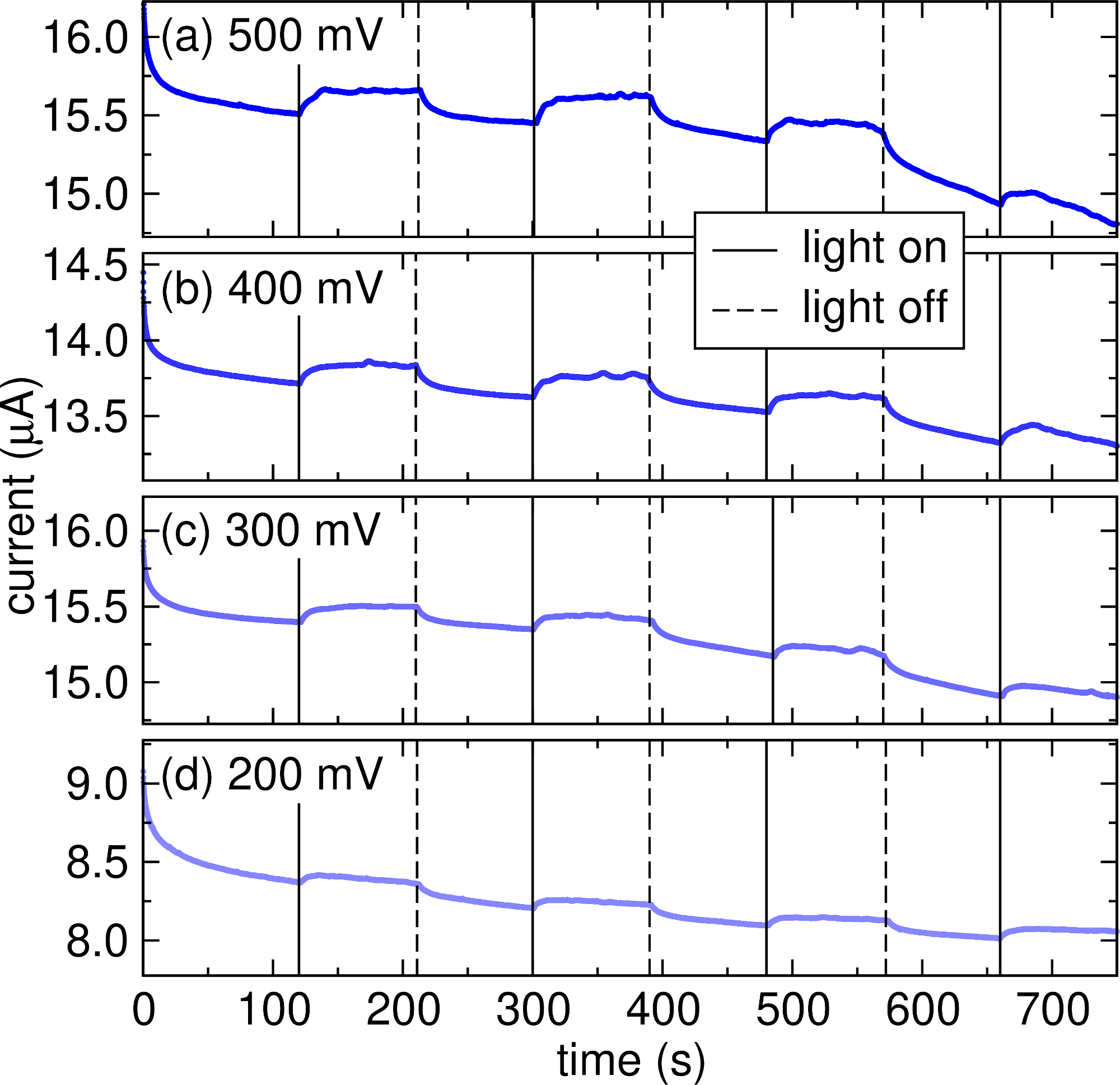}
\caption{\textbf{DC Polarization experiments for LLTO upon 350 nm irradiation using non-blocking Li electrodes.} A potential hold of (a) 500 mV, (b) 400 mV, (c) 300 mV, and (d) 200 mV were applied to measure the steady state current with and without irradiation in 90 second intervals, with one set shown by the arrows.}
\label{fig:350nmLiDC}
\end{figure}

\begin{figure}[hbt!]
\centering
\includegraphics[width=0.45\linewidth]{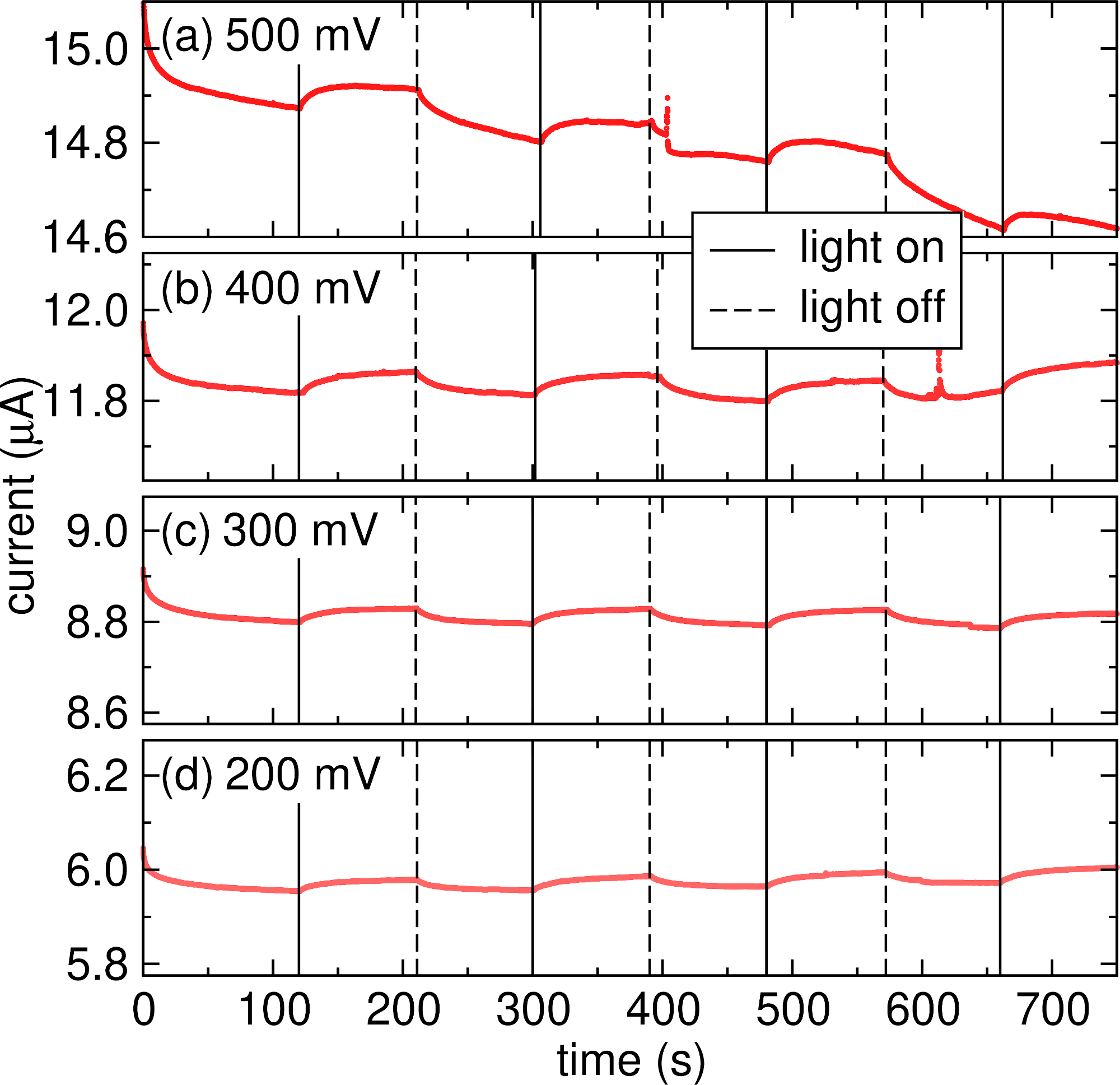}
\caption{\textbf{DC Polarization experiments for LLTO upon 800 nm irradiation using non-blocking Li electrodes.} A potential hold of (a) 500 mV, (b) 400 mV, (c) 300 mV, and (d) 200 mV were applied to measure the steady state current with and without irradiation in 90 second intervals, with one set shown by the arrows.}
\label{fig:800nmLiDC}
\end{figure}
\begin{figure}[hbt!]
\centering
\includegraphics[width=0.45\linewidth]{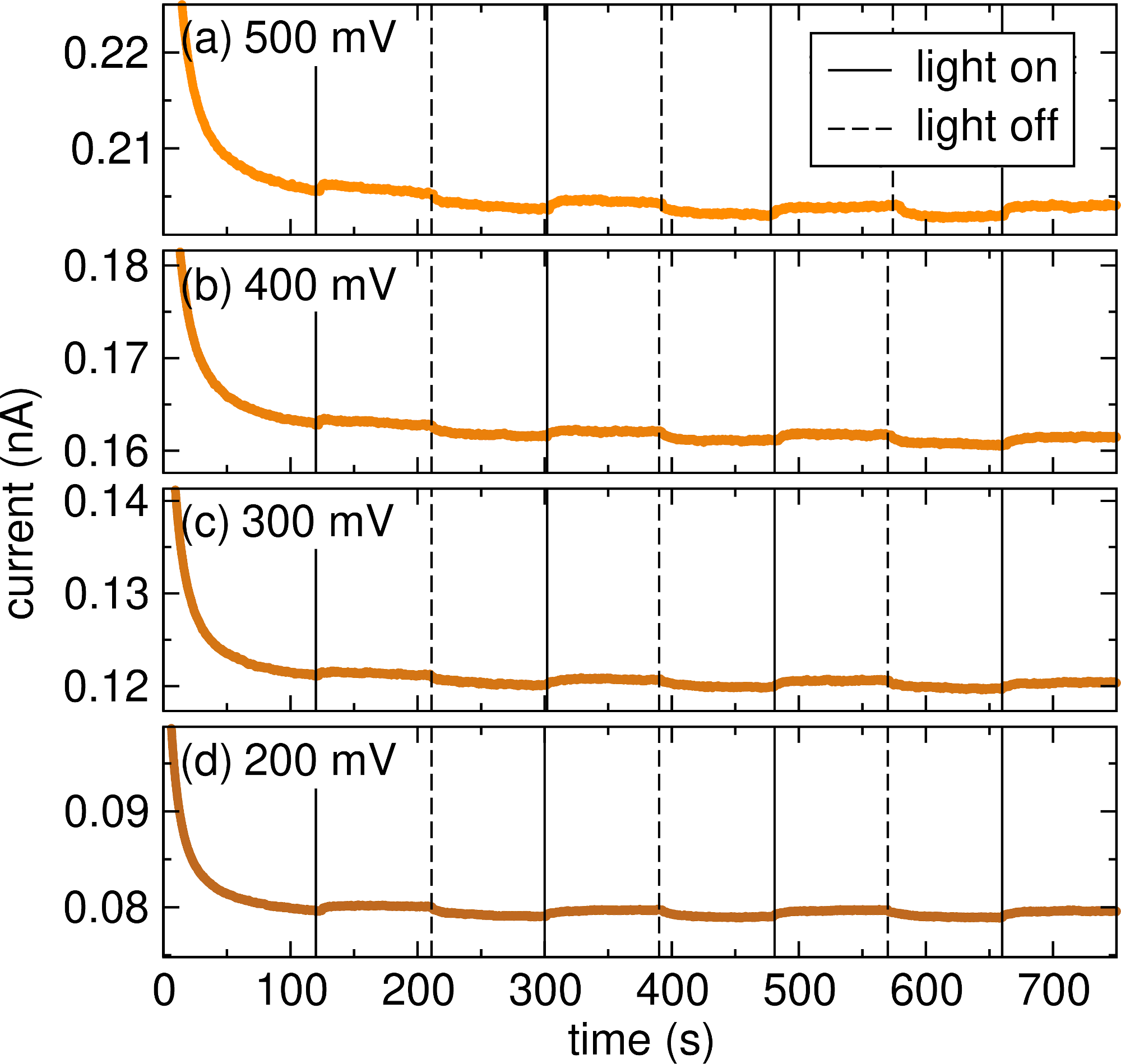}
\caption{\textbf{DC Polarization experiments for LLTO upon 350 nm irradiation using blocking Au electrodes.} A potential hold of (a) 500 mV, (b) 400 mV, (c) 300 mV, and (d) 200 mV were applied to measure the steady state current with and without irradiation in 90 second intervals, with one set shown by the arrows.}
\label{fig:350nmAuDC}
\end{figure}

\begin{figure}
\centering
\includegraphics[width=0.45\linewidth]{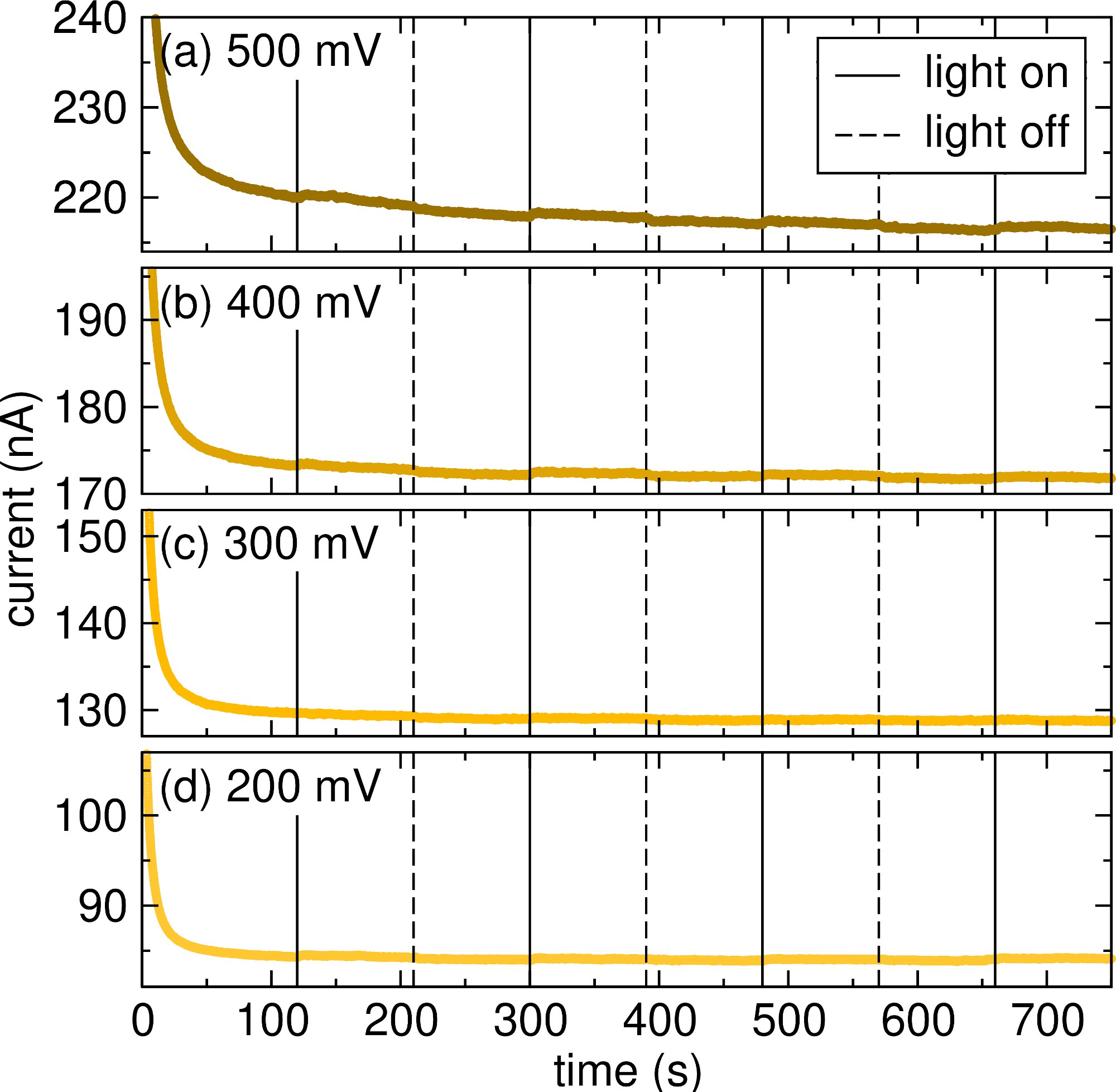}
\caption{\textbf{DC Polarization experiments for LLTO upon 800 nm irradiation using blocking Au electrodes.} A potential hold of (a) 500 mV, (b) 400 mV, (c) 300 mV, and (d) 200 mV were applied to measure the steady state current with and without irradiation in 90 second intervals, with one set shown by the arrows.}
\label{fig:800nmAuDC}
\end{figure}

\begin{figure}[hbt!]
    \centering
    \includegraphics[width=0.7\linewidth]{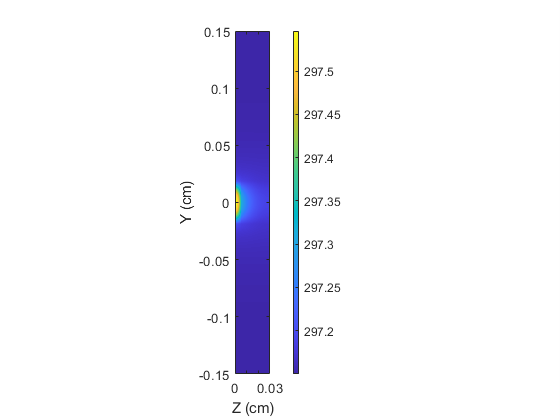}
    \caption{\textbf{Cross-section of the sample after 1 s of ultrafast laser heating, modeled with an 800 nm, 20 mW excitation with a penetration depth of \( 7\ \boldsymbol{\mu} \)m} The color bar corresponds to the temperature of the sample in Kelvin, with the initial temperature at 297 K. The sample and excitation is centered at x = 0 and y = 0.} 
    \label{fig:800slice}
\end{figure}

\begin{figure}[hbt!]
    \centering
    \includegraphics[width=0.7\linewidth]{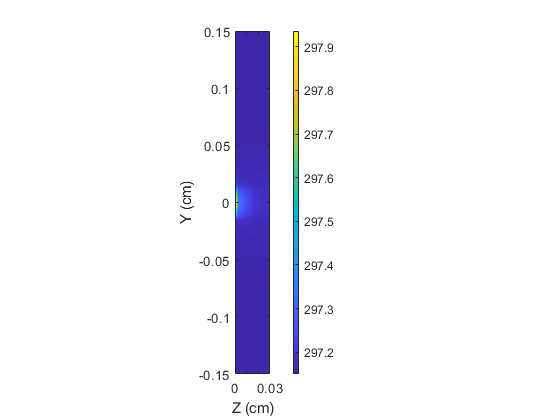}
    \caption{\textbf{Cross-section of the sample after 1 s of ultrafast laser heating, modeled with a 350 nm, 16 mW excitation with a penetration depth of 30 nm} The color bar corresponds to the temperature of the sample in Kelvin, with the initial temperature at 297 K. The sample and excitation is centered at x = 0 and y = 0.} 
    \label{fig:350slice}
\end{figure}

\begin{figure}[hbt!]
    \centering
    \includegraphics[width=0.7\linewidth]{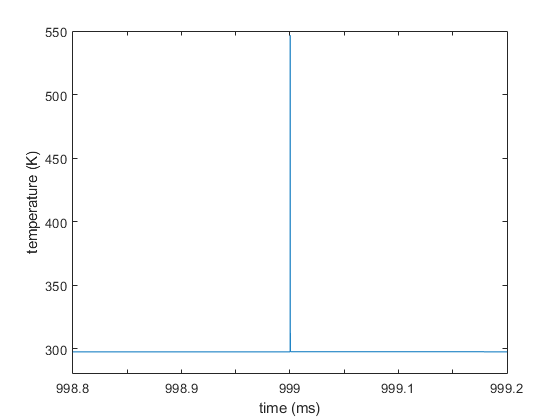}
    \caption{\textbf{The peak temperature in the sample during and after the final pulse for the 800 nm, 20 mW light simulation.} The nearly instantaneous rise and decay of heat occurs in less than 100 ns and indicative of the extremely fast response induced by an ultrafast laser pulse.} 
    \label{fig:800pulse}
\end{figure}

\begin{figure}[hbt!]
    \centering
    \includegraphics[width=0.7\linewidth]{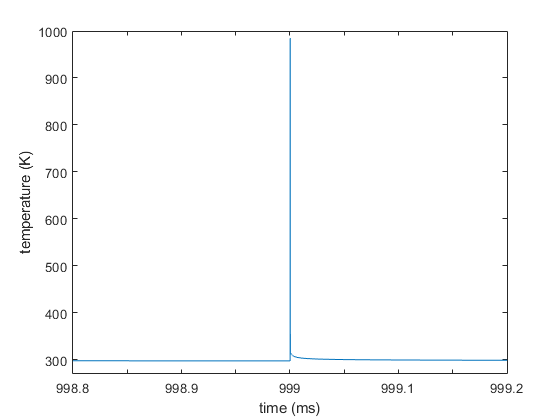}
    \caption{\textbf{The peak temperature in the sample during and after the final pulse for the 350 nm, 16 mW light simulation.} The nearly instantaneous rise and decay of heat occurs in less than 100 ns and indicative of the extremely fast response induced by an ultrafast laser pulse.} 
    \label{fig:350pulse}
\end{figure}

\begin{figure}[hbt!]
    \centering
    \includegraphics[width=0.7\linewidth]{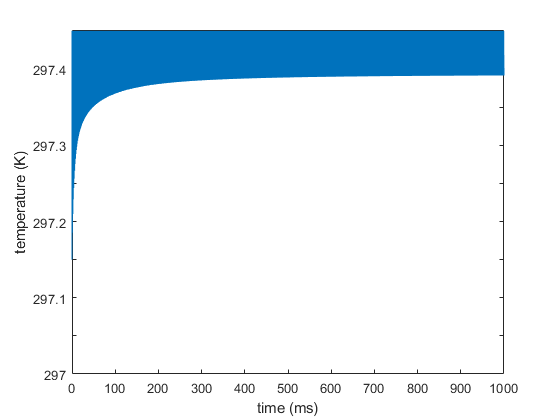}
    \caption{\textbf{The baseline maximum temperature in the sample for 800 nm, 20 mW light.} The simulation demonstrates that the sample's baseline temperature reaches steady state within one second. Each pulse still causes a nearly instantaneous rise and decay, but a steady baseline temperature of 297.55 K is reached in several tenths of a second.} 
    \label{fig:800base}
\end{figure}

\begin{figure}[hbt!]
    \centering
    \includegraphics[width=0.7\linewidth]{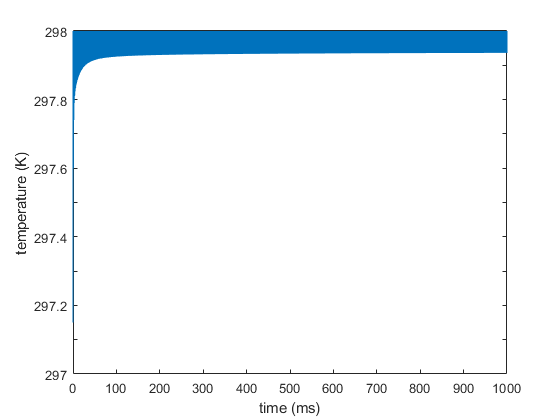}
     \caption{\textbf{The baseline maximum temperature in the sample for 350 nm, 16 mW light.} The simulation demonstrates that the sample's baseline temperature reaches steady state within one second. Each pulse still causes a nearly instantaneous rise and decay, but a steady baseline temperature of 297.94 K is reached in several tenths of a second.} 
    \label{fig:350base}
\end{figure}

\begin{figure}[hbt!]
    \centering
    \includegraphics[width=1\linewidth]{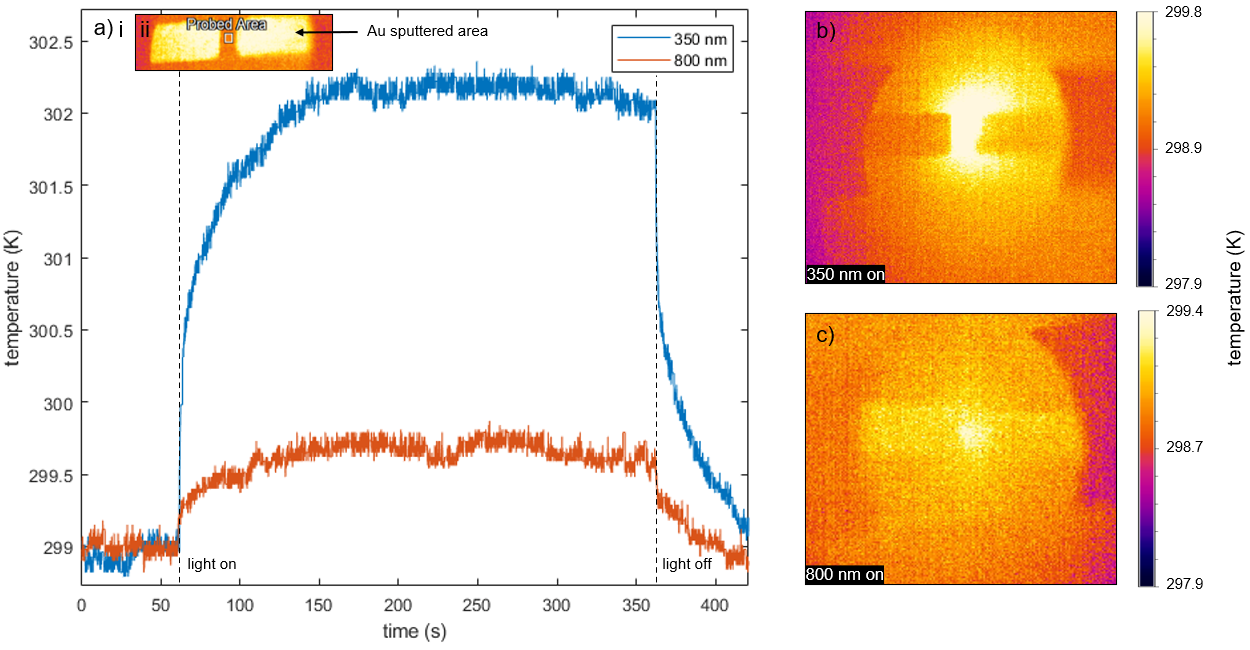}
    \caption{\textbf{Experimental measurement of the LLTO sample surface temperature with sputtered Au contacts under 800 nm, 20 mW and 350 nm, 16 mW illumination using an infrared thermal gun.} (a) Measured surface temperature of the probed sample area upon illumination between 60 and 360 seconds (i) and inset of the IR camera image, showing the probed area of the sample (ii). (b) Surface map of sample under 350 nm excitation and 800 nm excitation (c) focused between the Au electrodes.}
    \label{fig:irgun}
\end{figure}

\begin{figure}[hbt!]
    \centering
    \includegraphics[width=0.7\linewidth]{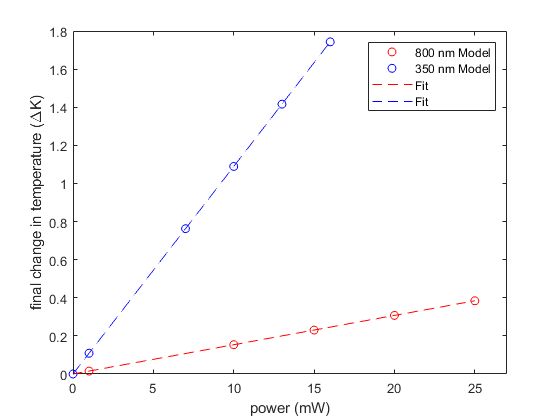}
     \caption{\textbf{The final temperature of the sample due to optical heating as a function of average 800 nm and 350 nm power.} The final temperature is determined by taking the average maximum temperature at all time points during and after the final pulse. Our results suggest that the 800 nm excitation raises the sample temperature by 0.03 K increase per mW of power and 0.11 K per mW for 350 nm.} 
    \label{fig:heatingmodellinregg}
\end{figure}

\begin{figure}[hbt!]
\centering
\includegraphics[width=0.45\linewidth]{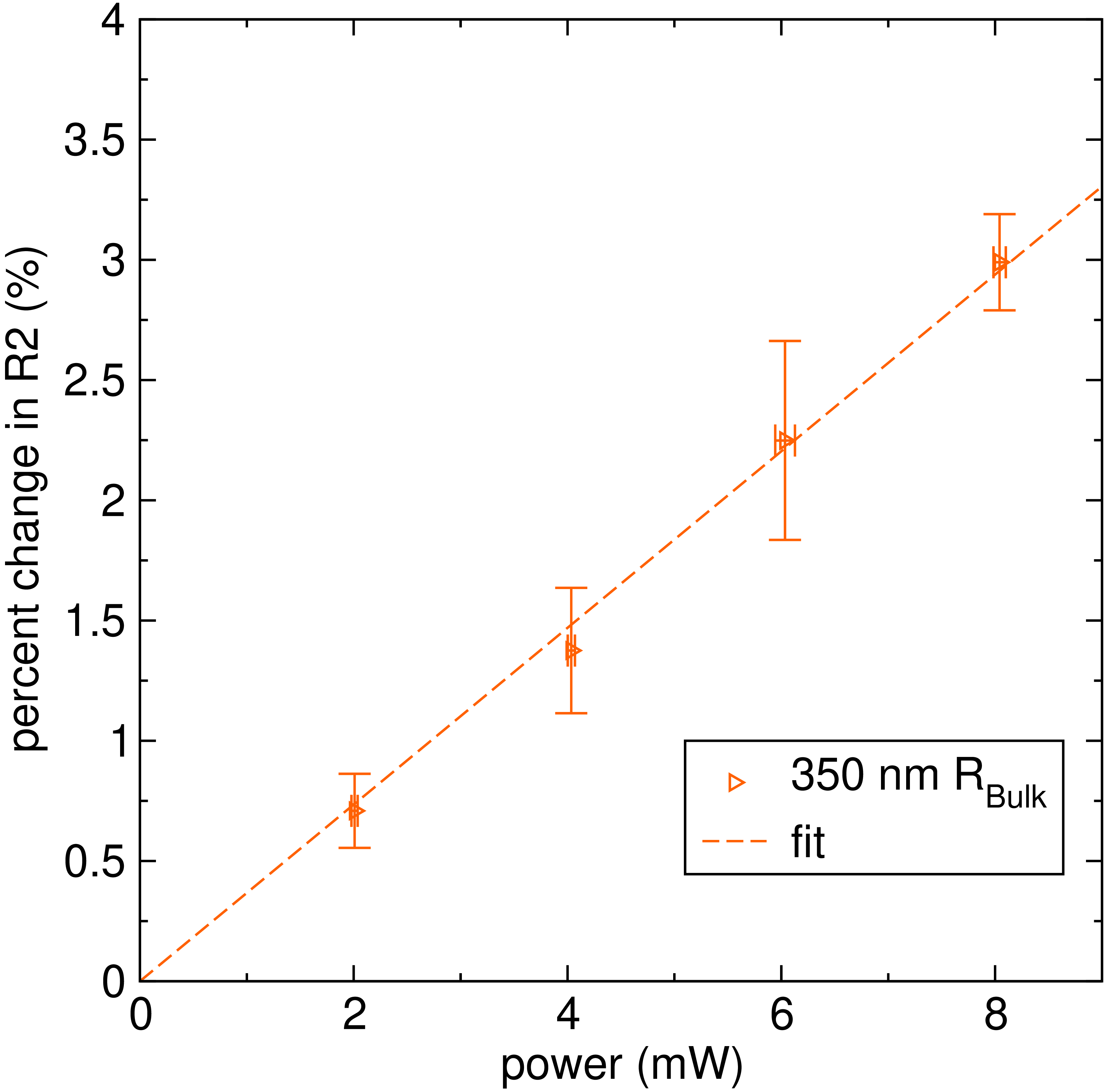}
\caption{\textbf{Percent change in $\mathbf{R}_{\mathbf{bulk}}$ upon irradiation of a 350 nm excitation source (1 kHz repetition rate, fs pulsed).} The $R^2$ is 0.9991 for the $R_{bulk}$. 1 mW of power causes a  0.37\% change in the fitted resistance.}
\label{fig:350nmlinregg}
\end{figure}

\clearpage{}
\section{Appendix}
\subsection{Consideration of alternate explanations for long lived photoinduced states}
\label{secA1}

It should be noted that explanations for long-lasting photoinduced states and dynamics involving electron, ion, and crystal lattice couplings include irreversible phase changes and light-induced material decomposition  \cite{zhao_quantification_2017}, reversible phase changes  \cite{gottesman_photoinduced_2015,marshall_probing_2021,shank_evidence_2015}, direct energy transfer to migrating ions  \cite{shank_evidence_2015,shishiyanu_mechanism_2011}, and dipole orientations or lattice distortions  \cite{wu_composition-dependent_2015, juarez-perez_photoinduced_2014}. Dipole orientations have the potential to screen charges, \cite{gottesman_photoinduced_2015,wu_composition-dependent_2015,harada_light-dependent_2022}, thereby enhancing ion mobility. In all cases, the dynamics that are observed are unequivocally too slow to be assigned purely to linear electronic effects, \cite{gottesman_photoinduced_2015,dequilettes_photo-induced_2016,juarez-perez_photoinduced_2014} considering that charge injection times can range from nanoseconds to microseconds \cite{marshall_probing_2021} while the migration of vacancies and defects occur up to microseconds, and in some cases for vacancies up to minute time-scales  \cite{azpiroz_defect_2015}. 

The 10’s of seconds timescale required to return to the pre-illumination impedance in LLTO is similar to the relaxation timescales measured by Hoke \textit{et al} with \ce{(CH_3NH_3)Pb(Br_{x}I_{1-x})_3} upon removal of white light, measured by photo-luminescence spectroscopy. \cite{hoke_reversible_2014} In that study, the minute timescale for relaxation was attributed to photo-driven halide segregation into I-rich and Br-rich domains, corroborated by X-ray diffraction. In more extreme cases, even longer relaxation timescales in dark measurements can last up to several hours like that measured by Gottesman \textit{et al} for \ce{CH_3NH_3PbBrI_3} under white light. \cite{gottesman_photoinduced_2015} Using Raman spectroscopy, the formation of new peaks associated with the liberation and migration of \ce{CH_3NH_3} (MA) cations , or i.e. suggesting that a more ordered, photo-driven structural transformation occurred to accommodate the alignment or enhanced rotation of MA cations. Interestingly in both scenarios of the inorganic-organic perovskites, full reversibility was confirmed as well as the lack of material degradation. 

The potentially convoluted effects of direct energy transfer to migrating ions must also be considered. The mechanism of direct energy transfer via light proposed by Shishiyanu \textit{et al.}\ portrays a simplified picture of how light energy can be decomposed into a thermal and quantum component as shown in Equation \ref{eq:3}, where $D$ is the diffusion coefficient, $D_0$ is the pre-factor term, $E_D$ is the activation energy, $k$ is the Boltzmann constant, and $T$ is the temperature, $E_T$ and $E_{hv}$ are the thermal and quantum parts of energy respectively, and $\eta$ is the efficiency of light radiation: \cite{shishiyanu_mechanism_2011}

\begin{equation} \label{eq:3}
D(T,hv) = D_0exp(-[E_D-\frac{\eta(E_{hv}-E_g(T))}{kT})
\end{equation}

 Equation \ref{eq:3} assumes that all absorbed optical energy goes into the ion hopping, which is not always the case. However, Equation \ref{eq:3} still broadly suggests the important effect of thermal and quantum energy transfer to ion migration. 

The equation for ionic conductivity $\sigma_{ion}$ is shown in Equation \ref{eq:4} where $\sigma_0$ is the Arrhenius pre-factor, $T$ is temperature, $E_a$ is the activation energy, and $k_B$ is the Boltzmann constant: 

\begin{equation} \label{eq:4}
\sigma_{ion}=\frac{\sigma_0}{T}exp(-E_a/k_BT) 
\end{equation}

When the pre-factor $\sigma_0$ in Equation \ref{eq:4} is expanded, we see that $\sigma_{ion}$ also depends on the entropy of migration $\Delta S_m$ and the enthalpy of migration $\Delta H_m$: \cite{muy_phononion_2021} 

\begin{equation} \label{eq:5}
\sigma_{ion} \propto exp(\Delta S_m/k_B)*exp(\Delta H_m/k_BT)*\frac{1}{T} 
\end{equation}

The value of $\Delta S_m$ is determined by collective phonon-ion interactions which are predicted to correlate with and enhance ionic conductivity \cite{vineyard_frequency_1957}. The value of $\Delta H_m$ is experimentally measured as $E_a$. If the band gap excitation is truly causing lattice distortions or dipole orientations, the increase in $\Delta S_m$ would consequently increase the ionic conductivity which we confirmed previously by driving phonon modes with THz fields in LLTO \cite{pham_correlated_2024}. The reorientation of dipoles can be confirmed with X-ray diffraction, as was shown for \ce{MAPbI_3}, where the ratio of two 2$\Theta$ angles corresponding to the change in rotation angle along the c-axis of the octahedra are measured and shown to change upon illumination. \cite{gottesman_photoinduced_2015} However, again, in the case of our time-resolved synchrotron data, only a thermal expansion is measured.

Finally, the 10’s of seconds timescale required to return to the pre-illumination impedance in LLTO is similar to the relaxation timescales measured by Hoke \textit{et al.}\ with \ce{(CH_3NH_3)Pb(Br_{x}I_{1-x})_3} upon removal of white light, measured by photo-luminescence spectroscopy. \cite{hoke_reversible_2014} In that study, the minute timescale for relaxation is attributed to photo-driven halide segregation into I-rich and Br-rich domains, corroborated by X-ray diffraction. In more extreme cases, even longer relaxation timescales in dark measurements can last up to several hours like that measured by Gottesman \textit{et al.}\ for \ce{CH_3NH_3PbBrI_3} under white light. \cite{gottesman_photoinduced_2015}. Considering that the observed relaxation time scales are shorter than minutes time scales and that a major photo-induced structure or phase change is absent according to time-resolved synchrotron XRD (Figure \ref{figsynchro-us}), the slightly faster timescales may indicate a more reversible local change like the slight \ce{TiO6} distortions predicted to occur at the saddle point of the migration pathway. Unfortunately, such small photo-induced distortions are difficult to confirm experimentally which renders this hypothesis inconclusive. A more rigorous molecular dynamics simulation could investigate this phenomena further, though it is beyond the scope of our study.

\subsection{Optical electrochemical cell}
\label{secA2}

The custom optical cell was designed to enable air-free, DC   experiments under simultaneous optical excitation, specifically to prevent the reaction of Li-metal during the measurement. The excitation enters through the centered hole of the nylon body bottom and the desired optical window, then to the un-sputtered area of the sample. The optical window is chosen to maximize transmission. An uncoated, fused silica window was used that transmits 80.0\% and 88.3\% of the 350 nm and 800 nm light respectively. 

\begin{figure}[hbt!]
\vspace{10pt}
\centering
\includegraphics[width=.8\linewidth]{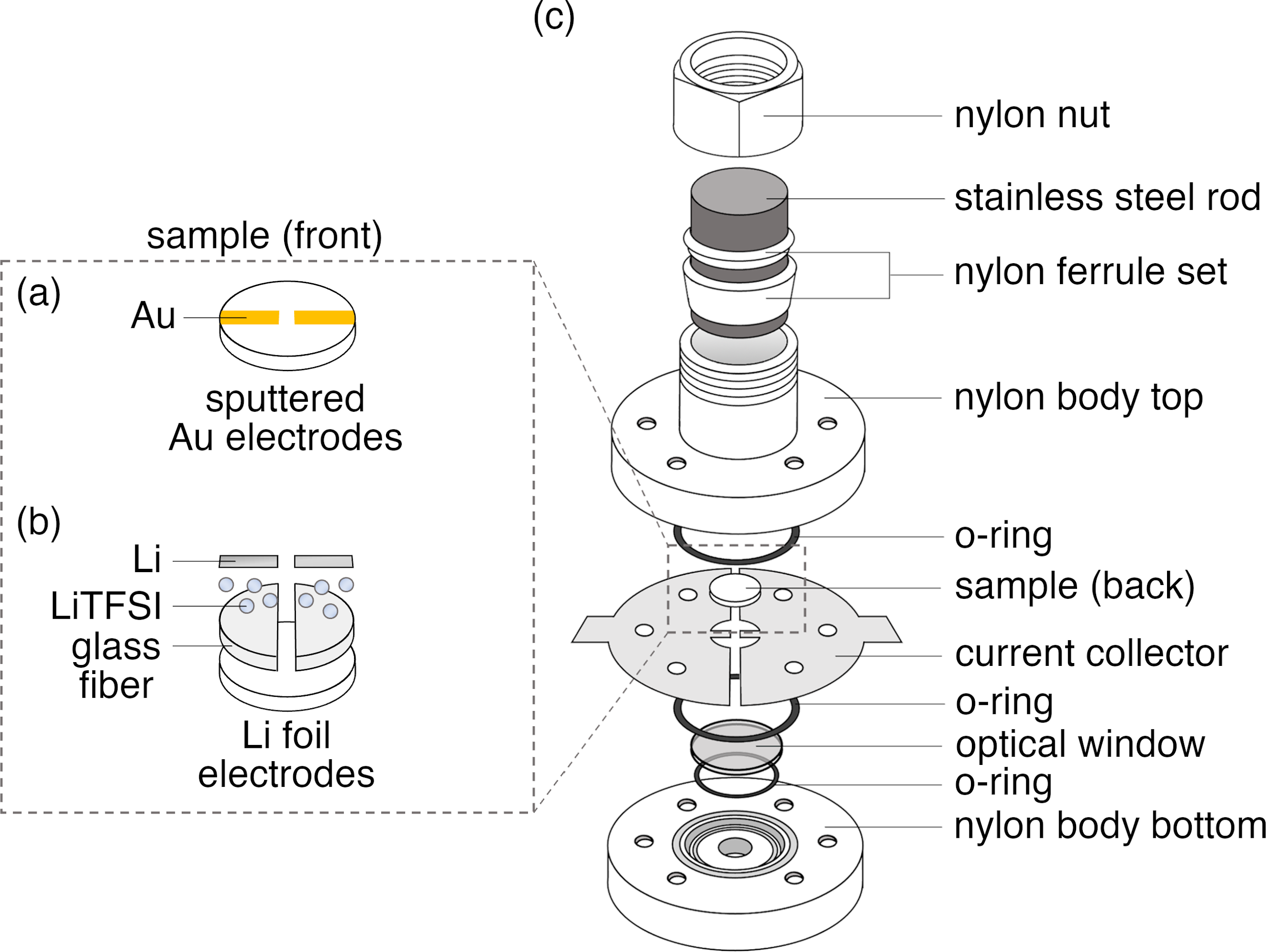}
\caption{\textbf{Blocking and non-blocking electrode geometry and corresponding custom electrochemical optical cell design.} (a) Blocking electrode geometry utilizing Au sputtered electrodes on the plane of the sample. (b) Non-blocking electrode set up utilizing Li foil and a glass fiber separator soaked in 1 M LiTFSI. (c) Optical cell design used to conduct electrochemical measurements under simultaneous photo-excitation with the set-up described in (b). Six screws are used to assemble the cell through the six corresponding, depicted holes shown in both nylon body parts.}
\vspace{10pt}
\label{fig:opticalcell}
\index{figures}
\end{figure}

The stainless steel current collector was machined to match the in-plane electrode geometry of the cell. The stainless steel rod is used to apply pressure to the sample and ensure contact between the electrodes and the current collector. The nylon nut and ferrule set is used to secure the top barrel section of the nylon body top, fully enclosing the cell against air exposure. Six screws are used to enclose the cell by screwing into the threaded holes of both nylon body parts. 

\subsection{Heating electrochemical cell}
\label{secA3}

\begin{figure}[hbt!]
\centering
\includegraphics[width=.6\linewidth]{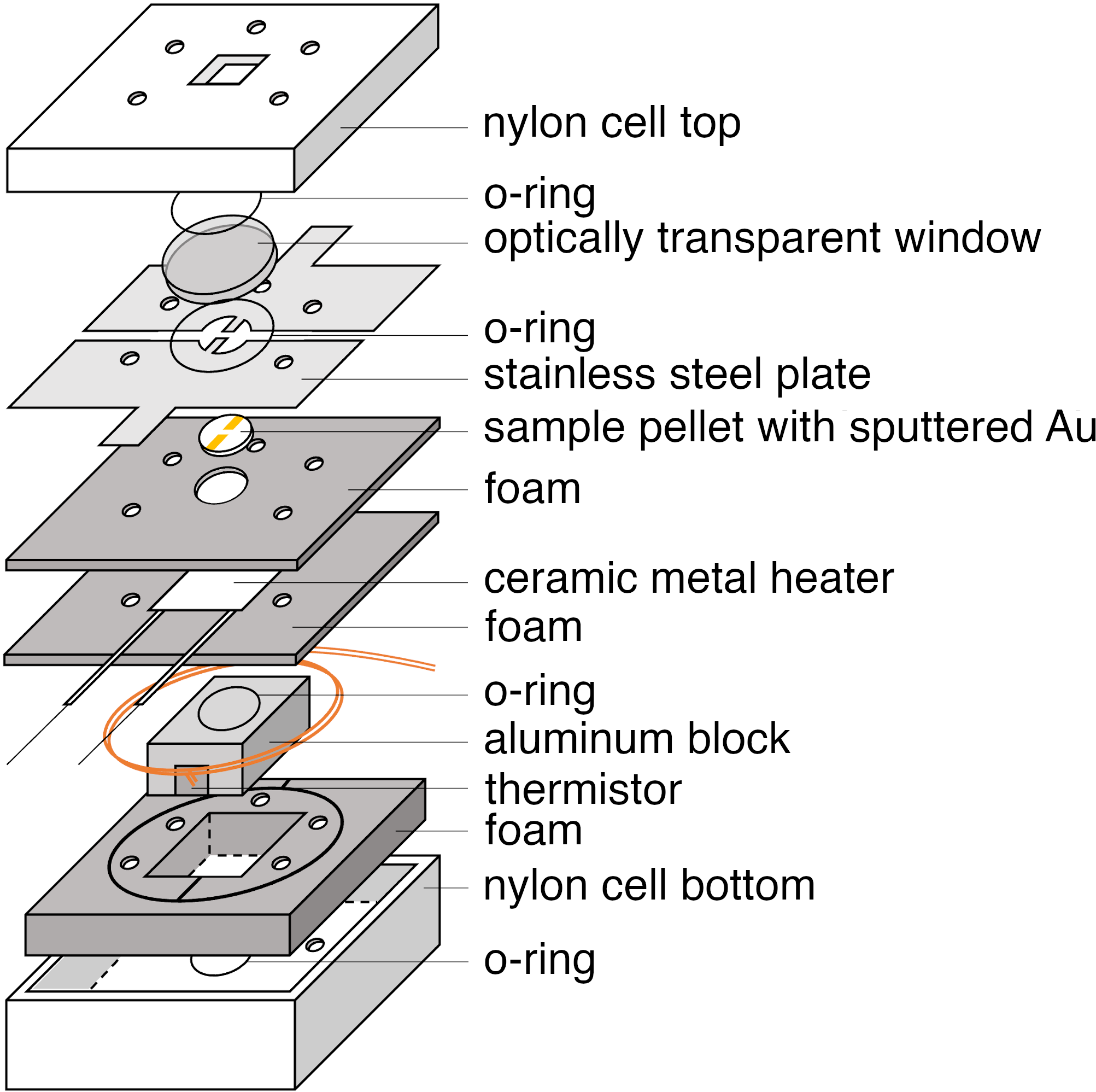}
\caption{\textbf{Custom electrochemical heating cell design.} Electrochemical heating cell set-up to obtain the power-to-temperature calibration curve and collect EIS data below 32 MHz with the 1260A Solartron. The cell components are compressed and held together with screws that fit through the five holes depicted. Each screw is secured with wingnuts. \textit{Reprinted from Pham et al Rev. Sci. Instrum. 95, \textbf{2024}, 073004.}}
\label{fig:heatingcell}
\index{figures}
\end{figure}

The heating cell was designed to conduct electrochemical experiments with an in-plane sample geometry like that shown for the optical cell, but with an additional temperature controller. The heating cell was also used to conduct DC measurements for the Au-electrode set up described in Figure \ref{fig:opticalcell}a. To insulate the cell, chemical-resistant nylon foam sheets were used. A  20.0 mm x 20.0 mm Thorlabs HT24-24 W Metal Ceramic Heater was used in conjunction with the externally powered TC-48-20 OEM to control the internal temperature of the cell via the thermistor. An optically transparent window can be used to enable simultaneous photo-excitation, or for non-air sensitive samples, a Teflon window with a centered drilled through hole can be used to fill the gap between the sample and nylon cell top. The nylon cell top contains two o-ring wells to seal the stainless steel plate, or current collector, and the optically transparent window. Five screws and wingnuts are used to secure the cell through the five depicted non-threaded holes in Figure \ref{fig:heatingcell}.

\clearpage

\newpage
\bibliography{references}

\end{document}